\PassOptionsToPackage{pdfpagelabels=false}{hyperref}
\documentclass[fleqn,usenatbib]{mnras}

\usepackage[T1]{fontenc}
\usepackage{ae,aecompl}

\usepackage{graphicx}	\usepackage{amsmath}	\usepackage{amssymb}	\usepackage{color}

\def \g3c {G$^{3}$C\,}
\def \mvir {M_{200}}
\def \rvir {R_{200}} 
\newcommand{\nfof}{N_{\text{FoF}}}

\newcommand{\logmstar}{\log[M_\star/(h^{-1} M_\odot)]}
\newcommand{\logmhalo}{\log[M_{200}/(h^{-1} M_\odot)]} \newcommand{\logmh}{\log[M_{\text{halo}}/(h^{-1}M_\odot)]}
\newcommand{\eagle}{EAGLE}\newcommand{\gama}{GAMA}\newcommand{\gamamock}{GAMA-Mock}\newcommand{\sdss}{SDSS}

\title[Stellar Mass Segregation in Galaxy Groups]
{Galaxy And Mass Assembly (GAMA): The absence of stellar mass segregation in galaxy groups and 
consistent predictions from GALFORM and EAGLE simulations}

\author[P. R. Kafle et al.]
{P. R. Kafle,$^{1}$ \thanks{E-mail: prajwal.kafle@uwa.edu.au, prrajkafle@gmail.com}
A. S. G. Robotham,$^{1}$
C. del P. Lagos,$^{1}$ 
L. J. Davies,$^{1}$
A. J. Moffett,$^{1}$
\newauthor 
S. P. Driver,$^{1,2}$
S. K. Andrews,$^{1}$
I. K. Baldry,$^{3}$
J. Bland-Hawthorn,$^{4}$
S. Brough,$^{5}$
\newauthor
L. Cortese,$^{1}$
M. J. Drinkwater,$^{6}$ 
R. Finnegan,$^{1}$
A. M. Hopkins$^{5}$ 
and J. Loveday $^{7}$
\newauthor
\\
{\footnotesize $^{1}$ ICRAR, The University of Western Australia,35 Stirling Highway, Crawley, WA 6009, Australia}\\
{\footnotesize $^{2}$ SUPA, School of Physics \& Astronomy, University of St Andrews, North Haugh, St Andrews KY16 9SS, UK}\\
{\footnotesize $^{3}$ Astrophysics Research Institute, Liverpool John Moores University, IC2, Liverpool Science Park, 146 Brownlow Hill, Liverpool L3 5RF, UK}\\
{\footnotesize $^{4}$ Sydney Institute for Astronomy, School of Physics A28, University of Sydney, NSW 2006, Australia}\\
{\footnotesize $^{5}$ Australian Astronomical Observatory, PO Box 915, North Ryde, NSW 1670, Australia}\\
{\footnotesize $^{6}$ Department of Physics, The University of Queensland, Brisbane, QLD 4072, Australia}\\
{\footnotesize $^{7}$ Astronomy Centre, University of Sussex, Falmer, Brighton BN1 9QH, UK}\\
}

\pubyear{2016}
\begin{document}
\label{firstpage}
\pagerange{\pageref{firstpage}--\pageref{lastpage}}
\maketitle

\begin{abstract}
We investigate the contentious issue of the presence, or lack thereof, of satellites mass segregation 
in galaxy groups using the Galaxy And Mass Assembly (GAMA) survey,  the GALFORM semi-analytic and  the EAGLE 
cosmological hydrodynamical simulation catalogues of galaxy groups. We select groups with halo mass 
$12 \leqslant \log(M_{\text{halo}}/h^{-1}M_\odot) <14.5$ and redshift $z \leqslant 0.32$ and probe the 
radial distribution of stellar mass out to twice the group virial radius. All the samples are carefully 
constructed to be complete in stellar mass at each redshift range and efforts are made to regularise the analysis 
for all the data. Our study shows negligible mass segregation in galaxy group environments with absolute gradients 
of $\lesssim0.08$ dex and also shows a lack of any redshift evolution. Moreover, we find that our results at least 
for the GAMA data are robust to different halo mass and group centre estimates. Furthermore, the EAGLE data allows 
us to probe much fainter luminosities ($r$-band magnitude of 22) as well as investigate the three-dimensional 
spatial distribution with intrinsic halo properties, beyond what the current observational data can offer. 
In both cases we find that the fainter EAGLE data show a very mild spatial mass segregation at $z \leqslant 0.22$, 
which is again not apparent at higher redshift. 
Interestingly, our results are in contrast to some earlier findings using the Sloan Digital Sky Survey. 
We investigate the source of the disagreement and suggest that subtle differences between the group finding algorithms could be the root cause.
\end{abstract}


\section{Introduction}
Both theoretical modelling of galaxy formation and observations reveal
that most of the stellar material in the Universe resides in groups of a few $10^{12} M_{\sun}$ and larger masses   
\citep[e.g.][etc]{1958ApJS....3..211A,1977ApJ...211..311R,1982ApJ...255..382H,1982ApJ...257..423H,1983ApJS...52...61G,
2000ARA&A..38..289M,2004MNRAS.348..866E,2006ApJS..167....1B,2007ApJ...671..153Y,2009ApJ...697.1842K,2011MNRAS.416.2640R,2013MNRAS.436..380N,
2014A&A...566A...1T,2014MNRAS.441.1270L,2015arXiv151105856S}.
Moreover, it is known that galaxies residing in a group environment follow a very different evolutionary course 
compared to that of isolated systems \citep{1974Natur.252..111E,1984ApJ...281...95P}.
Therefore, the group environment is clearly an important factor in understanding both structure formation and galaxy evolution
at intermediate local mass densities. 

In current galaxy formation models, galaxies in the groups can be broadly classified in 
two categories: central galaxies and satellite galaxies \citep[e.g.][etc]{2005ApJ...633..791Z,2011MNRAS.410..417S}.
Central galaxies are located near the centre of a parent dark matter halo.
Under the current paradigm of hierarchical structure formation, 
the central galaxies of the subhalo that gets accreted to the dominant nearby halo are called satellites. 
Subsequently the accreted galaxies (satellites) are potentially quenched 
by environmental effects, such as gas stripping by ram-pressure \citep{1972ApJ...176....1G,2009MNRAS.399.2221B}, 
removal or reduction of hot/cold gas or even the stellar components of the satellite galaxy due 
to tidal stripping \citep{1996Natur.379..613M,2006PASP..118..517B}.  
Thus, to develop a viable theory of galaxy formation 
it is important to understand the processes that could influence the abundance 
and distribution of satellites in galaxy groups. 

A spatial distribution of stellar mass segregation in any dynamical system, 
ranging from globular clusters to galaxy groups and clusters, is an important indicator of their evolutionary history and dynamical friction time-scales. 
The sinking of heavier objects in a gravitational potential well of stellar \citep{1998MNRAS.295..691B} 
and galaxy \citep{1977MNRAS.179...33W,2004MNRAS.352L...1G,2005ApJ...619..193M} clusters has been repeatedly observed. 
Broadly, the mass segregation is known to be either primordial \citep{1997MNRAS.285..201B}, 
meaning clusters may form with the most massive galaxies concentrated near the centre, or dynamical \citep{Allison2009} caused by migration of the most 
massive galaxies into the centre of the cluster via relaxation. 
If dynamical friction in the group environment plays a dominant role,
then the effect on the stellar mass distribution in galaxy groups should be detectable. 
Conversely, if there is an absence of spatial mass segregation 
in groups, it could possibly mean that the contribution of ongoing star formation in 
galaxies, or tidal stripping of satellite galaxies as they fall inward, or 
that the group is continually fed by new merging groups in a dominant process directing the distribution of the mass in groups. 
In other words, it means that the relaxation time of the galaxy groups is significantly longer than their crossing time.

With the advent of large redshift surveys it has only recently become possible to study mass segregation in galaxy groups 
in great detail using the Sloan Digital Sky Survey (SDSS; \citealt{2000AJ....120.1579Y}) and zCOSMOS \citep{2012ApJ...753..121K}.  
Recently, \cite{2015MNRAS.448L...1R} showed the presence of mass segregation trends in SDSS, meaning 
satellites of higher masses are systematically concentrated 
close to the group-centre at all halo mass ranges. 
This is in close agreement with earlier studies using different data sets, for example, 
\citet[SDSS]{2008arXiv0805.0002V} and \citet[zCOSMOS]{2012A&A...539A..55P}.
Similarly, \cite{2014MNRAS.443.2679B} also find some mass segregation, but at small group radii of $\lesssim 0.1$ times the virial radius.
Simultaneously, there are also evidence to contradict the existence of mass segregations in galaxy groups.
For example, \cite{2013MNRAS.434.3089Z} fail to observe strong mass segregation in X-ray selected groups up to $z\sim1.7$.
However, they could not rule out that this might be due to a bias introduced by their sample selection. 
Similarly \cite{2012MNRAS.424..232W}, using galaxy group catalogues created from SDSS DR7, with a
modified implementation of the group-finding algorithm in \cite{2007ApJ...671..153Y}, also find 
no evidence of mass segregation for satellites at any halo mass range.

Despite this large body of work, there is little consensus on the presence or the strength of mass segregation in galaxy groups. 
On the theory side there have been analogous studies \citep[e.g.][etc]{2004MNRAS.348..333D,2005MNRAS.359.1537R,2016MNRAS.455..158V} that show
the segregation of dark matter subhaloes in numerical simulations of various extents, 
but also see \cite{2004MNRAS.352..535D,2008MNRAS.391.1685S,2009ApJ...692..931L} for contradictory findings.
In the future, it would be valuable to combine the theoretical work with the studies of satellites mass segregation in galaxy groups 
to better understand the galaxy-halo connection and the various physical processes, such as how galaxies populate haloes.

In this work we aim to resolve the contentious issue of the presence or absence of mass segregation in galaxy groups.
For this we investigate group catalogues from three types of data, observed: using the Galaxy and Mass Assembly survey 
(\gama; \citealt{2011MNRAS.413..971D,2015MNRAS.452.2087L}); 
semi-analytics: using the \gama\ lightcone mock catalogues 
(\gamamock; \citealt{2013MNRAS.429..556M} using the \cite{2014MNRAS.439..264G} variant of 
the \textsc{galform} semi-analytic model of galaxy formation \citep{2000MNRAS.319..168C,2015arXiv150908473L}),
and cosmological hydrodynamical simulation: using the Evolution and Assembly of 
GaLaxies and their Environments (\eagle; \citealt{2015MNRAS.446..521S,2015arXiv151001320M}).
In order to make the results from all the three data sets comparable, 
we homogenise the estimates of physical quantities such as group-centric distance, stellar mass and halo virial properties. 

Throughout the paper we assume a cosmological constant $\Omega_\Lambda=0.75$, matter density $\Omega_M=0.25$ and 
$h=H_0/(100$ kms$^{-1}$ Mpc$^{-1})$.
Also, $\log$ stands for logarithm to the base 10, and $r$ and $R$ represent the spherical (3D) and projected (2D) radii respectively. 
For conciseness, we use the standard notation $($ and $]$ to denote open and closed intervals respectively. 

This paper is arranged as follows. 
In Section~\ref{sec:data}, we describe \gama, \gamamock\ and \eagle\ data, 
their corresponding group catalogues and the derivation of quantities relevant to our analysis. 
In Section~\ref{sec:result} we present our main results.
In Section~\ref{sec:discussion} we provide a detailed comparison
of our work with the available group catalogues of Sloan Digital Sky Survey (SDSS) data
and also among different group catalogues of SDSS. 
Our findings are summarised in Section~\ref{sec:conclusion}.

\section{Data}\label{sec:data}
We use data from three main sources. 
We first describe the data sets individually followed by how we compute   
informations relevant to the study of mass segregation within galaxy groups. 
 
\subsection{Galaxy and Mass Assembly (GAMA)}\label{sec:gama}
The \gama\ survey is a spectroscopic and multiwavelength survey of galaxies carried out on the Anglo-Australian Telescope \citep{2011MNRAS.413..971D,2015MNRAS.452.2087L}. 
Details of the \gama\ survey characteristics are given in \cite{2011MNRAS.413..971D}, 
with the survey input catalogue described in  \cite{2010MNRAS.404...86B}, the spectroscopic processing outlined in
\cite{2013MNRAS.430.2047H}, and the spectroscopic tiling algorithm explained in \cite{2010PASA...27...76R}. 
The survey has obtained ~300,000 galaxy redshifts to $r<19.8$ mag over $\sim 286$ deg$^2$,
with the survey design aimed at providing uniform spatial completeness \citep{2010MNRAS.404...86B,2011MNRAS.413..971D}.
Here we use the complete northern equatorial sample referred to as \gama-II-N covering over three $12\times5$ deg$^2$ fields centred at $9^{\text{h}}$(G09), 
$12^{\text{h}}$(G12) and $14.5^{\text{h}}$(G15) RA and approximately $0^\circ$ declination, described in full in \cite{2015MNRAS.452.2087L}.

The data used here primarily focusses on the \gama\ galaxy groups, which are constructed
using an adaptive Friends-of-Friends (FoF) algorithm, linking galaxies in projected and line-of-sight separations. 
For the full details about the algorithm, diagnostic tests, construction and caveats of the group catalogue 
we refer the reader to \cite{2011MNRAS.416.2640R}.

\subsection{Semi-analytic Data (GAMA-Mock)}\label{sec:sam}
We use the \gama\ light cone mock catalogues constructed from the \textsc{galform} semi-analytic model of galaxy formation \citep{2014MNRAS.439..264G}.
The model uses analytic, physically motivated equations to follow the evolution of the baryonic components of galaxies (stars, cold gas, hot gas, and their metals).
\textsc{galform} makes use of these equations to populate dark matter halo merger trees that are generated from 
$N$-body simulations (a new Millennium Simulation MS-W7; \citealt{2013MNRAS.428.1351G} ) of dark matter.
The MS-W7 simulation uses $2160^{3}$ particles, each with a mass of $9.35\times 10^{8}h^{-1}$ M$_\odot$ in a box of side $500 h^{-1}$ Mpc 
(see \citealt{2005Natur.435..629S}, for details of the original Millennium Simulation).

\textsc{galform} models the following processes in galaxies: (i) the collapse and merging of dark matter haloes, (ii)
gas heating and cooling through shocks and radiative cooling inside dark matter haloes, leading to the formation of galactic disks, (iii) quiescent star formation in galactic disks,
(iv) supernovae and AGN feedback from the photo-ionization of the intergalactic medium, (v) chemical enrichment of gas and stars, (vi) galaxy mergers leading to the formation of stellar spheroids,
which can also trigger a starburst, and (vii) the collapse of gravitationally unstable disks, which also leads to the formation of spheroids and starbursts.
The scale size of the disk and bulge of galaxies is also computed.
The galaxy luminosities are determined by combining the star formation and metal enrichment histories with stellar population synthesis models for each galaxy.
The attenuation of starlight by dust is included based on radiative transfer calculations.
The final product of the calculation is a prediction of the number and properties of galaxies that reside within dark matter haloes of different masses. 
The model we use here is that of \cite{2014MNRAS.439..264G}.
The outputs of the model are placed in a lightcone using the technique described in \cite{2013MNRAS.429..556M}, 
and the details for how the \gama\ selection and sky areas were applied to the lightcones are described in \cite{2015MNRAS.454.2120F}. 

Importantly, the construction of the group catalogue for \gamamock\ and estimates of the group properties, e.g.,
galaxy stellar mass, group centre and projected distance etc are done similar to the \gama\ data.
This effort is to ensure consistency and make \gama\ and \gamamock\ results comparable. 
However, \cite{2011MNRAS.416.2640R} showed that the main discrepancy between the observed \gama\ group catalogue 
and the mock lightcone is that there is a relative excess of very
compact groups in the mocks data otherwise there is a high degree of agreement 
between the two data sets.

\subsection{EAGLE: a cosmological hydrodynamical simulation }\label{sec:eagle}
The final data set that we compare to our observational results is taken from the 
Evolution and Assembly of GaLaxies and their Environments (\eagle, \citealt{2015MNRAS.446..521S}) simulation.
\eagle\ is a suite of cosmological hydrodynamical simulations performed at a range of numerical resolutions, 
in periodic volumes with a range of sizes, and using a variety of subgrid implementations to model physical
processes below the resolution limit. 
One of the unique aspects of \eagle\ is 
the plethora of sub-grid baryonic physics included in the model: (i) radiative cooling and photoheating rates, 
(ii) star formation, (iii) stellar evolution and metal enrichment, (iv) stellar feedback, and (v) black hole growth and AGN feedback. 
These physical models are the key to reproducing a large set of properties of the observed galaxy population in the local Universe. 
For more details of the simulation we refer the reader to \citet{2015MNRAS.446..521S}.   
The subgrid parameters of the \eagle\ reference model are calibrated 
to the $z=0$ galaxy stellar mass function, stellar mass-black hole mass relation, and stellar 
mass-size relations (see \citealt{2015MNRAS.450.1937C} for details and motivation).
The \eagle\ reference model reproduces many observed galaxy 
relations that were not part of the calibration set, 
such as the evolution of the galaxy stellar mass function \citep{2015MNRAS.450.4486F}, 
galaxy sizes (\citealt{2015arXiv151005645F}), optical colours \citep{2015MNRAS.452.2879T}, 
and atomic \citep{2016MNRAS.456.1115B} and molecular gas content \citep{2015MNRAS.452.3815L}, 
amongst others. Thus it is an excellent test-bed to compare with our observations.

We use the public database of \eagle\ described in \cite{2015arXiv151001320M}. 
In particular, we focused our attention on the reference model of \eagle\ run in a cubic volume of $100^3$ Mpc$^3$ on a side 
with $2\times 1504^3$ dark matter and gas particles (particle masses are of $9.7\times 10^6$ and $1.81\times 10^6$ in $M_\odot$ units, 
respectively, and a physical resolution of $0.7$ kpc). 

The full phase space information of the galaxies and the halos hosting them are provided in the simulated data. 
However, to facilitate comparison with the observations, we transform the given phase space information into projected space. 
The first step for this is to compute apparent redshift for each galaxy given its cosmological redshift 
$(z_{\text{snapshot}})$ and peculiar velocity ($\mathbfit{v}$).  
We take the cartesian-z direction, with unit vector $\hat{e}_z$, as the direction of line-of-sight.
The formula for the apparent redshift is given by,

\begin{equation}
\label{eqn:zapp}
z = ( 1 + z_{\text{snapshot}} ) (1 + \mathbfit{v}.\hat{e}_z/c) - 1,
\end{equation}
where c denotes the speed of light.

Besides the properties of individual galaxies, we must also estimate parameters for the groups within \eagle.
To be consistent with the observational galaxy group catalogue of \gama, 
we use the galaxy groups in \eagle\ found using a FoF method, 
where a linking length of 0.2 times the average inter-particle spacing has been assumed.
For more details about the group finding in the \eagle\ data refer to \cite{2015MNRAS.446..521S}.
The position of the centre of the gravitational potential well corresponds to the
position of the most bound particle in the group. 

\subsection{Intricacies of the data: deriving galaxy and group properties}\label{sec:derivedquant}
Here, we describe the derivation of the galaxy and group parameters relevant to our study in each of our data sets.
The galaxy properties needed for this work are stellar mass and projected distance from the central galaxy of the group.
Similarly, the group information required in our study are: the position of the group central galaxy, 
overall velocity dispersion of the galaxies, and virial mass and virial radius of the host halo. 
In some cases, those are already provided by the respective survey teams such as: 
central galaxy, halo virial mass, radius etc in \eagle; 
group centre, stellar mass in \gama\ or the \gamamock. 
However, to ensure consistency, where possible, 
we re-estimate the above quantities for all the three data sets using a common method
as described below:

\subsubsection{Galaxy stellar mass ($M_\star$):}\label{sec:logmstar}
We estimate the galaxy stellar mass for \gama\ and \gamamock\ using a colour-based relation  
\begin{multline}
\label{eqn:stellarmass}
\logmstar = -0.4i + 0.4 \mu(z) - \log(1+z) \\
         + (1.2117 -0.5893z) + (0.7106-0.1467z)(g-i) - 2 \log(h/0.7), 
\end{multline}
where $M_\star$ is the stellar mass expressed in the units of solar mass $M_\odot$, $z$ is the galaxy redshift, 
$g$ and $i$ are the observed \gama\ $g$ and $i$ band apparent Kron magnitudes and $\mu(z)$ is the luminosity distance modulus. 
Both $g$ and $i$ are in the observer's frame and thus, implicitly accounts for a k-correction as well as stellar population as a function of colour. 
The above formula is adopted from \cite{2015MNRAS.447.2857B} and is derived following the approach of \cite{2011MNRAS.418.1587T}. 
In the case of \eagle\ data we directly use the provided stellar mass values.
Note we do this to avoid applying uncertain k-correction terms to get the data in observed rather than the native rest-frame.

\subsubsection{Group occupancy ($\nfof$):}\label{sec:nfof}
In this work we utilise the latest version of the \gama\ galaxy group catalogue (\g3c v08). 
The FoF grouping parameters are tuned to the mock catalogues and were optimized for groups with $\nfof>4$,
where $\nfof$ is the number of members grouped together by the FoF algorithm. 
A visual inspection of the phase space (distance-velocity plane) of \gama\ groups confirms that groups with $\nfof \leqslant  4$ are more contaminated by 
interlopers \citep[][refer to the Fig.13 which shows the group quality as a function of $\nfof$]{2011MNRAS.416.2640R}, 
while member selection for groups with $\nfof>4$ is in better agreement with the expectation of a smooth distribution of galaxies with a maximum velocity that decreases with radius. 
We therefore restrict our study to \gama\ groups with $\nfof > 4$,
and impose the same limit on the \gamamock\ and \eagle\ groups as well. 

\subsubsection{Group centre and projected distance ($R$):}\label{sec:proj_radius} 
\cite{2011MNRAS.416.2640R} identify the group centre in each group using three definitions of group centre: the moments derived centre of light (Cen), 
an iterative method rejecting the galaxy farthest away from the centre of light (recalculated at each iteration) 
until one galaxy is remained (the `iterative' centre IterCen), and the brightest group galaxy (BGG).
All galaxies that are not central galaxies are classified as satellite galaxies. 
In most cases ($\sim90\%$) the iterative central galaxy coincides with the BGG, while the centre of light is more discrepant. 
\cite{2015MNRAS.452.3529V} perform a detailed analysis of the lensing signal of \gama\ groups comparing the different centre definitions and confirm the results of \cite{2011MNRAS.416.2640R}, 
that is the BGG and the iterative centre both represent the group centre to a good degree, while the centre of light poorly represents the group centre. 
We consider the brightest absolute r-band magnitude galaxy in groups as a proxy for the BGG and also as the central galaxy of the group as per \cite{2011MNRAS.416.2640R}.
However, we investigate the robustness of our results to different group centre definitions in Section~\ref{sec:cendef}.
Once identified, we exclude the central\footnote{Note, by construction the $\sim60\%$ of ungrouped galaxies in \gama\ are simply central galaxies of a halo, 
where \gama\ is not deep enough to observe any satellites.} galaxies from our analysis and only keep the satellites. 
The group centric distance $R$ (in units of $h^{-1}$ Mpc) is essentially a 
projected comoving distance separation of the satellite galaxy to the right ascension (RA) and declination (Dec) of the group centre.

\subsubsection{Group virial mass ($M_{200}$) and radius ($R_{200}$):}\label{sec:virialprop}
There are different ways to estimate the total dynamical mass of the host halo associated with galaxy groups. 
For example, using weak lensing \citep[e.g.][etc]{1996ApJ...466..623B,2002MNRAS.335..311G,2004AJ....127.2544S,
2005ApJ...634..806P,2015MNRAS.452.3529V,2015MNRAS.446.1356H},
from abundance matching \citep[e.g.][etc]{2010ApJ...717..379B,2010ApJ...710..903M,2013ApJ...770...57B,2013MNRAS.433..659H},
from halo occupation \citep[e.g.][etc]{2002ApJ...575..587B,2002MNRAS.329..246B,2005ApJ...631...41T},
from conditional luminosity function based modellings \citep[e.g.][etc]{2003MNRAS.339.1057Y,2006MNRAS.365..842C}, 
from the velocity dispersion of galaxy groups using the virial theorem \citep[e.g.][etc]{1997ApJ...476L...7C,2006eac..book.....S} etc. 
The virial mass ($M_{200}/(h^{-1} M_\odot)$) and the virial radius ($r_{200}/(h^{-1} {\text{Mpc}})$)\footnote{to convert spherical radius $r$ into projected radius $R$ and 
vice versa we use $r= \pi R/2$ \cite[Equation 6.26]{2006eac..book.....S}.}
for a given redshift $z$ are connected through the relation: 
\begin{equation}
\label{eqn:m200}
\mvir = \frac{4 \pi}{3} r_{200}^3 \Delta \rho_{\text{crit}}(z). 
\end{equation}
Here, we use the critical density $\rho_{\text{crit}} = 3 H^2(z)/(8 \pi \text{G})$, the halo average density is $\Delta=200$ 
times the $\rho_{\text{crit}}$ and the Hubble parameter as a function of redshift $H(z) = H_0 \sqrt{\Omega_M (1+z)^3 + \Omega_\Lambda}$ assuming
no curvature and a negligible radiation contribution.

To assign total halo masses to groups in our catalogues, we adopt the virial theorem based approach.
But we discuss the effects of using different halo mass measurements in our final results (Section~\ref{sec:discussion}).
Our virial measurements adopt the conventional definition, i.e., the virial radius $r_{200}$ is 
the radius in which the mean enclosed density is larger than $\Delta=200$ times the critical density at the 
respective redshift $\rho_{\text{crit}}(z)$. 
From the virial theorem we get
\begin{equation}
\label{eqn:r200}
\frac{\text{G} \mvir}{r_{200}} = (\sqrt{\alpha} \sigma_v)^2, 
\end{equation}
where the parameter $\alpha$ defines the nature of the overall velocity distribution of member galaxies in group.
Here, we assume $\alpha=3$ as suggested in, e.g., \citet{1997ApJ...485L..13C,2006eac..book.....S} etc,
which is valid for a case of isotropic velocity distribution. 
We estimate the group velocity dispersions ($\sigma_v$), 
using the technique known as the gapper-method presented in \cite{1990AJ....100...32B},
and also used in e.g. 2dFGRS Percolation Inferred Galaxy Group 
(2PIGG; \citealt{2004MNRAS.348..866E}), SDSS \citep{2007ApJ...671..153Y}, zCOSMOS \citep{2009ApJ...697.1842K}, 
\gama\ \citep{2011MNRAS.416.2640R} etc. 
Finally, solving Equations \ref{eqn:m200} and \ref{eqn:r200} simultaneously we obtain  
the values for both $\mvir$ and $\rvir$ for each galaxy group.
Again, the same method is used to measure virial properties of the group catalogues of all the three sets of data. 
\begin{figure*}
   \centering
   \includegraphics[width=2\columnwidth]{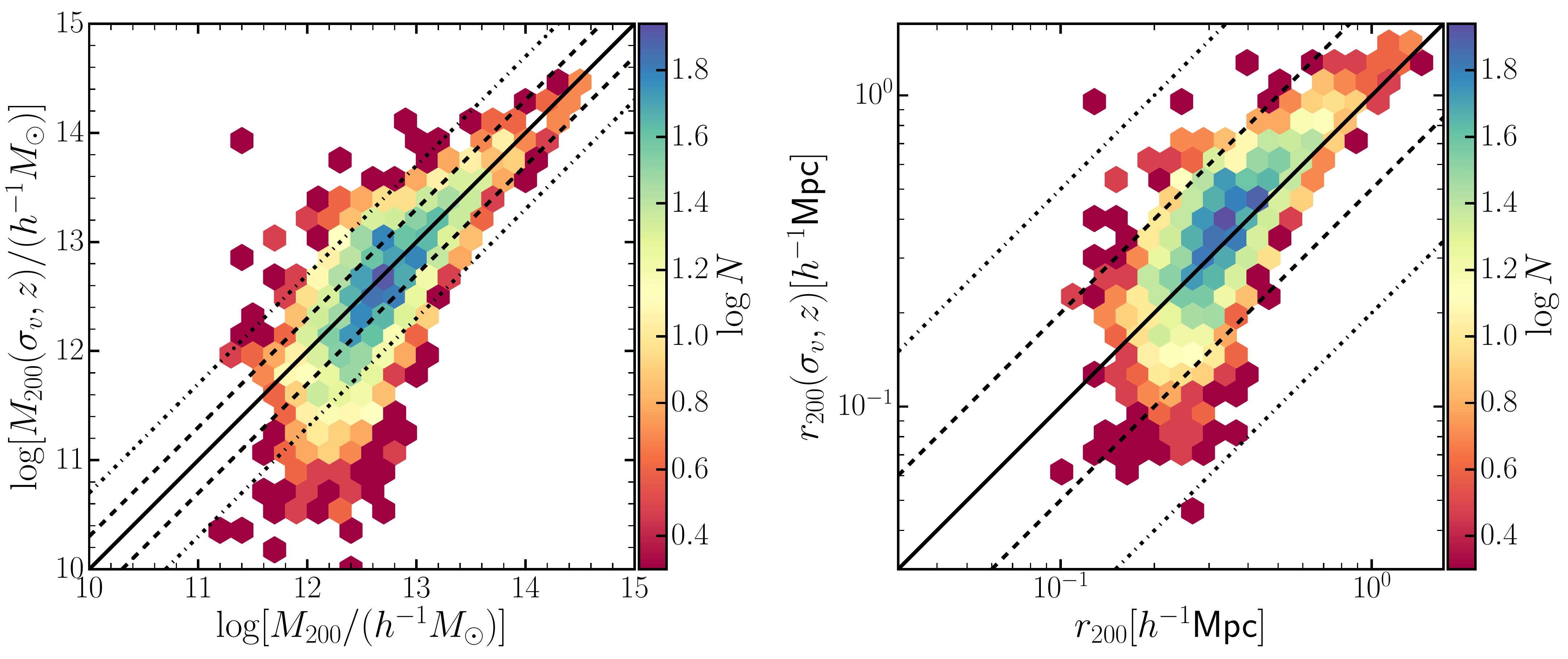}
   \caption{Comparison of intrinsic (along horizontal axis) and estimated (along vertical axis) virial properties of the \eagle\ galaxy groups.
            The panel on the left shows comparison of the halo masses $\log(\mvir)$ whereas panel on the right 
            compares the halo virial radius $r_{200}$.
            The solid, dashed and dotted lines are 1:1, 1:2, and 1:5 demarcation lines respectively.
            The colour of each pixel represents log number count of galaxies as labelled in the colour bar alongside.}
   \label{fig:halo_200}
\end{figure*} 

For the galaxy groups in the \eagle\ catalogue, $\mvir$ and $\rvir$ are already given/known. 
Thus we can compare our estimate of virial properties obtained 
using empirical method discussed above against the supplied values.
Figure~\ref{fig:halo_200} shows the comparison between the estimated virial properties,
i.e. $\mvir(\sigma_v,z)$ and $\rvir(\sigma_v,z)$ against the 
corresponding values intrinsically known from the \eagle\ simulations.
The solid, dashed and dotted lines in the figure represent the loci where the ratio of intrinsic and 
computed virial properties are 1:1, 1:2, and 1:5 respectively.
Both the $\mvir$ and $\rvir$ largely agree with each other at high masses and large radii. 
However, in the regions where $\rvir(\sigma_v, z)\lesssim0.1$ and $\log(\mvir(\sigma_v, z))\lesssim 11.5$ we can see significant deviation from the diagonal lines.
Note, the coloured pixels show number counts in a logarithmic scale.
As such, the number of discrepant groups at low $\mvir$ or $\rvir$ is small.
Nevertheless, there is a clear disagreement at low mass or radius. 
The disagreement is potentially due to a number of reasons.
For example, the underlying assumption in our estimates of $\rvir$ and $\mvir$ 
from Equations \ref{eqn:r200} and \ref{eqn:m200} 
is that the groups are virialised and are characterised by an isotropic distribution of velocities ($\alpha=3$), 
which may not necessarily be the case \citep{1993AJ....105.2035D}.
In case of anisotropy, $\alpha$ could be off by up to a factor of two \citep{1999ApJ...518...69M} resulting in systematic errors in our estimates. 
However, in our subsequent analysis we only use halos with $\log(\mvir)\geq12$, where 
the agreement between the estimated and intrinsic virial properties is reasonably good. 
Note, ideally the halos with intrinsic halo mass $\logmhalo\lesssim13$ and 
empirically measured halo mass $\log(\mvir(\sigma_v,z)) \lesssim12$
would have been included in our sample if the calibration given by Equation~\ref{eqn:r200} was perfect.
Since each halo is expected to be individually stellar mass limited at a given redshift, 
the effect of the missed halos due to poor halo-mass calibration will only be in the overall statistics in a given halo mass bin.
\begin{figure}
   \centering
   \includegraphics[page=1, width=0.95\columnwidth]{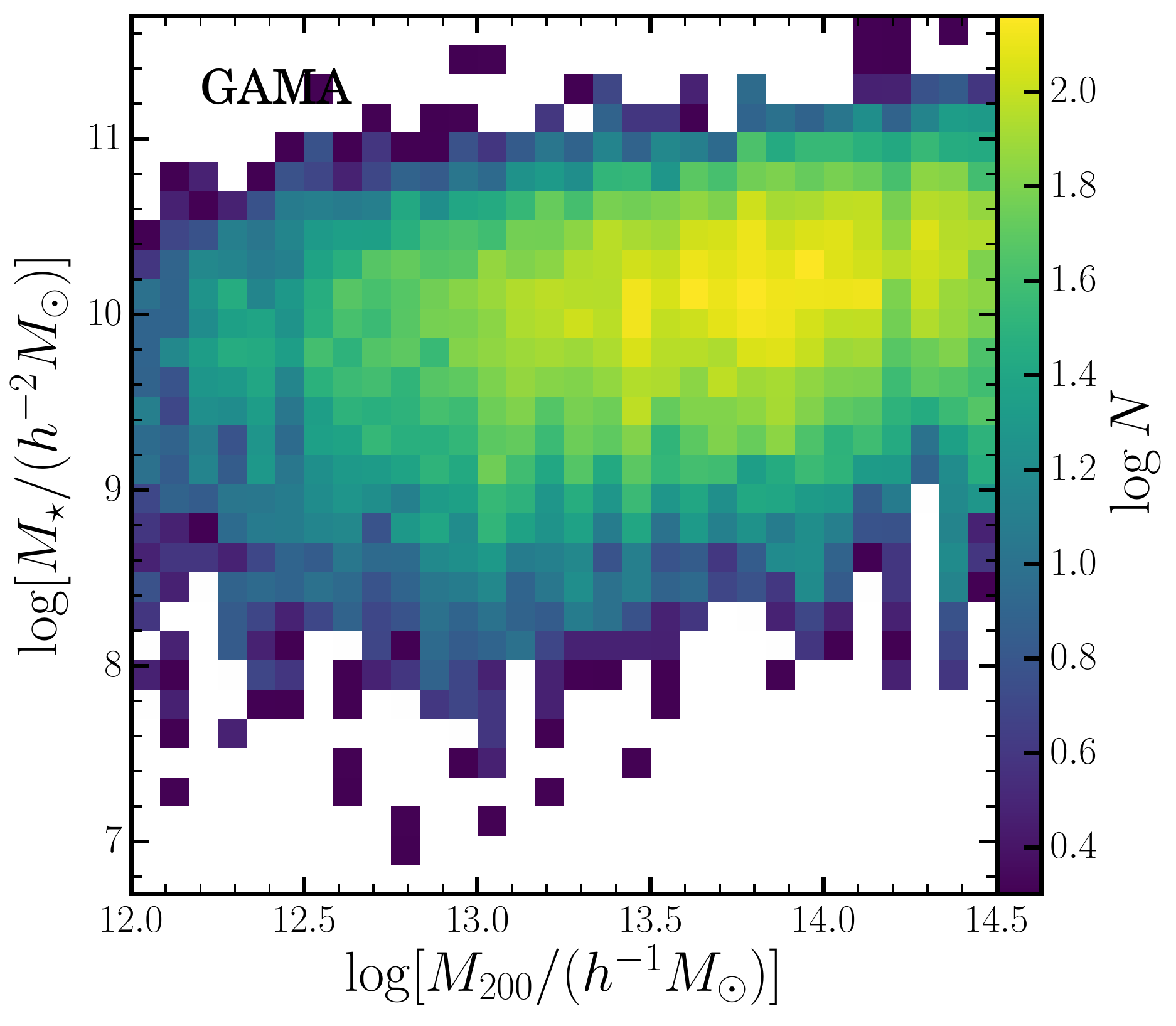}
   \includegraphics[page=1, width=0.95\columnwidth]{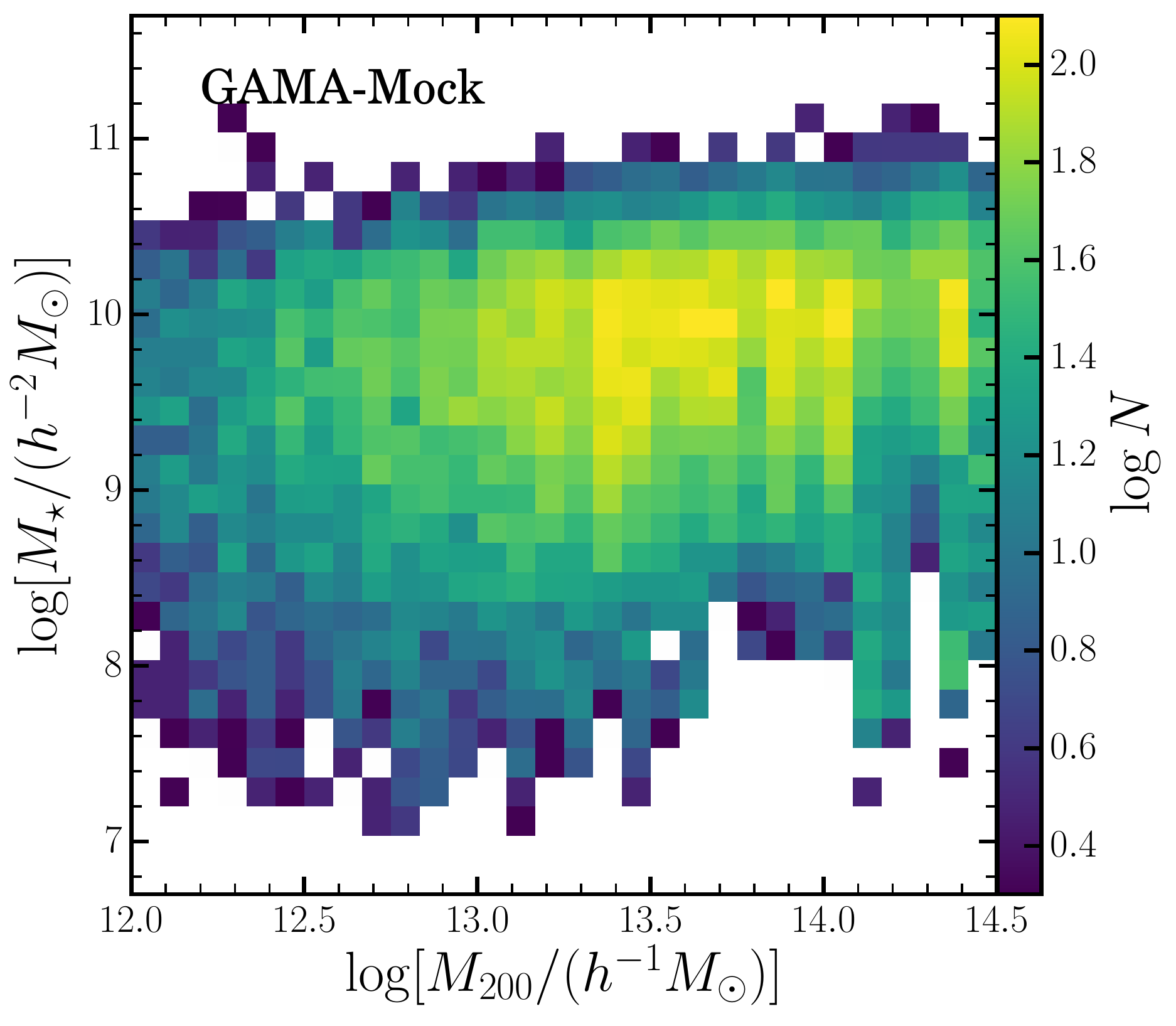}
   \includegraphics[page=1, width=0.95\columnwidth]{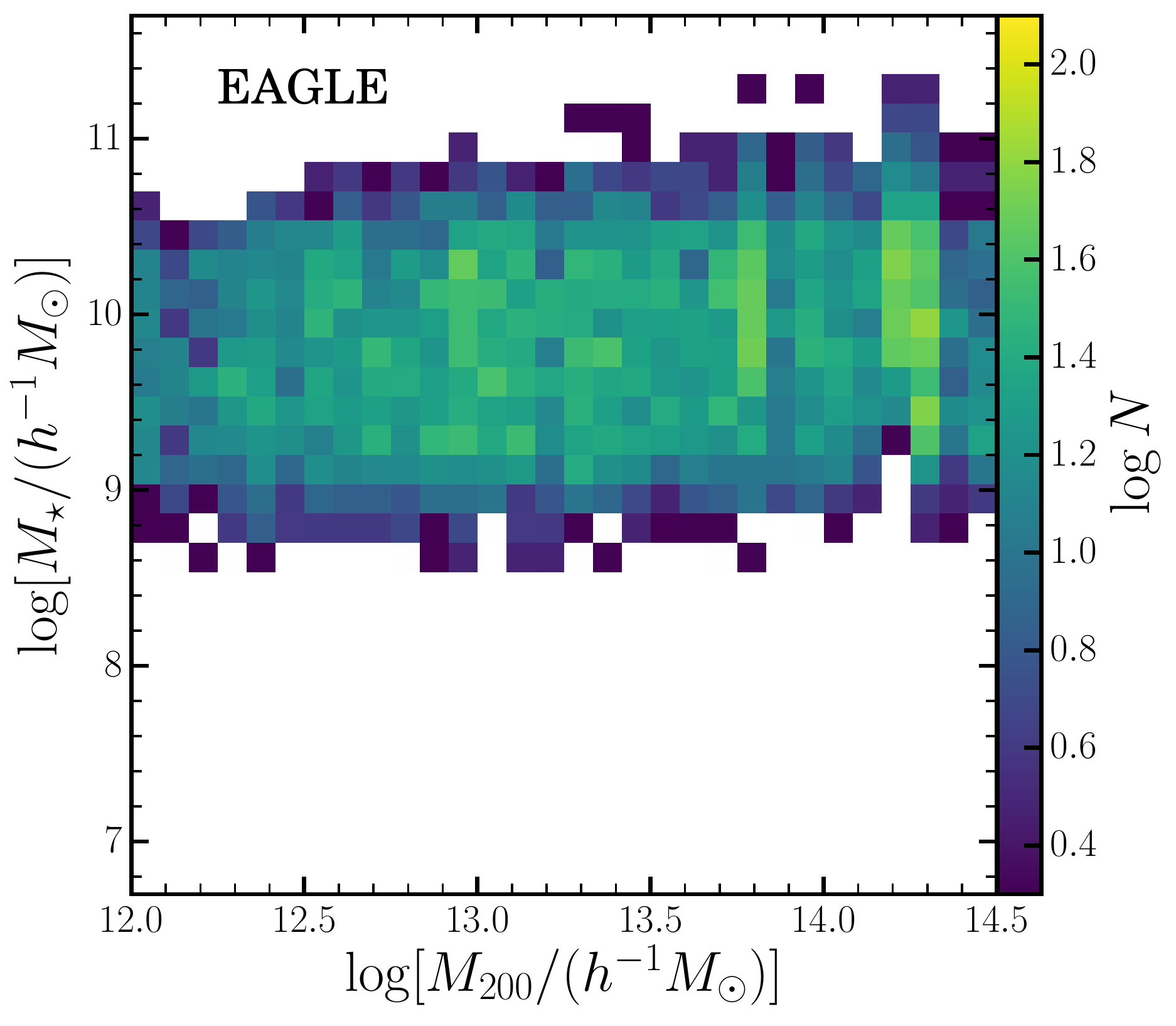}
   \caption{Stellar mass ($M_\star$) of the satellites versus the host halo mass $(\mvir$)
            as labelled in each panel from top to bottom is \gama, \gamamock\ and \eagle\ data.
            The colour of each pixel represents log number count of galaxies 
            as labelled in the colour bar alongside. 
            Note, this shows the entire sample for all the three data sets and is not just limited to the stellar mass complete sample.}
    \label{fig:mstar_halo}
\end{figure}

As a reference, in Figure~\ref{fig:mstar_halo} we show the joint distributions of the stellar mass 
($M_\star$) of the satellite galaxies in groups as a function of halo mass ($\mvir$) of the 
corresponding groups for all the three data sets (\gama, \gamamock\ and \eagle).
Note that the figure shows all the galaxies in groups, thus they may not be necessarily complete in stellar mass.
The figure is only presented to provide additional insight into the data, 
and also to guide the appropriate halo mass selection ranges that we use in this paper. 

\subsubsection{Volume limited samples}\label{sec:vollimit}
\begin{figure}
    \includegraphics[page=2, width=0.95\columnwidth]{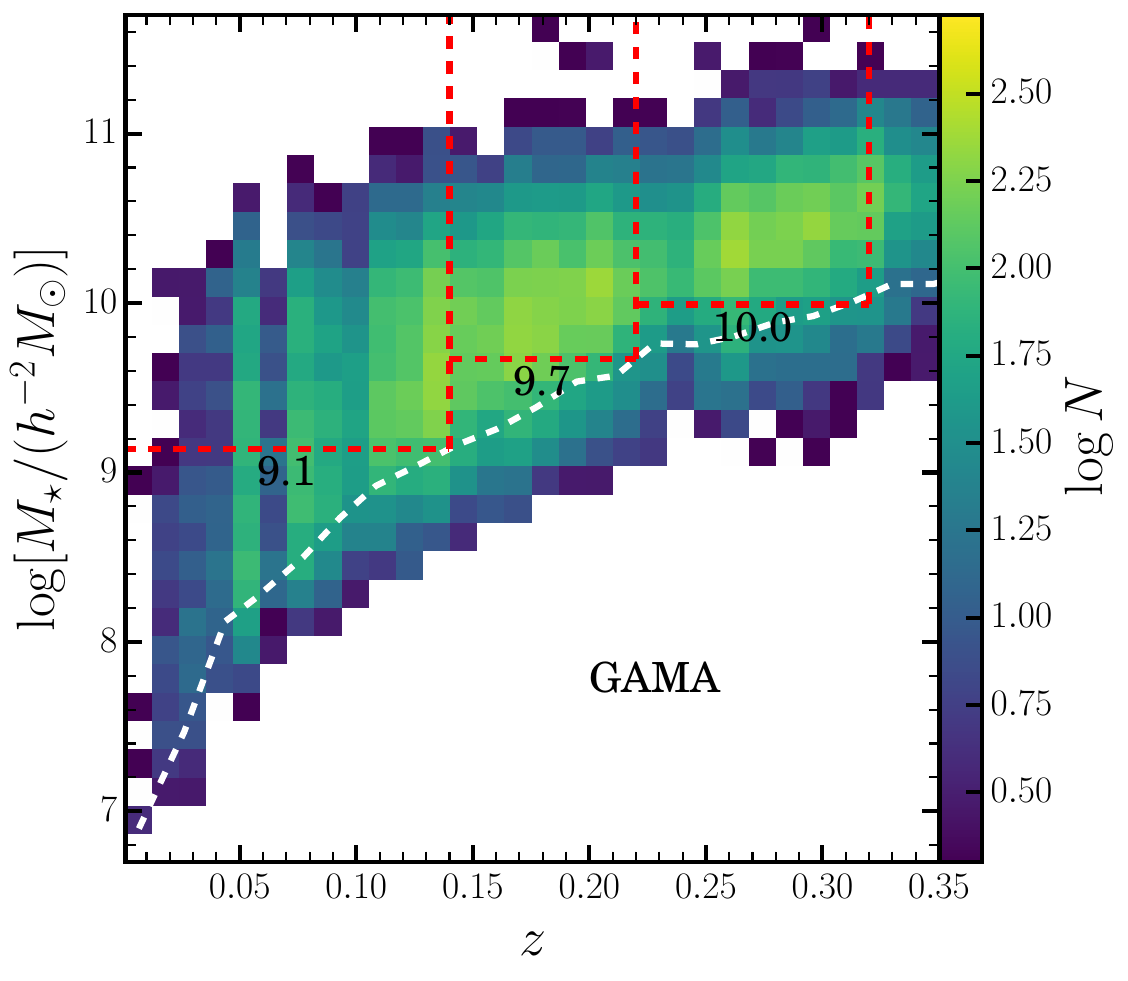}
    \includegraphics[page=2, width=0.95\columnwidth]{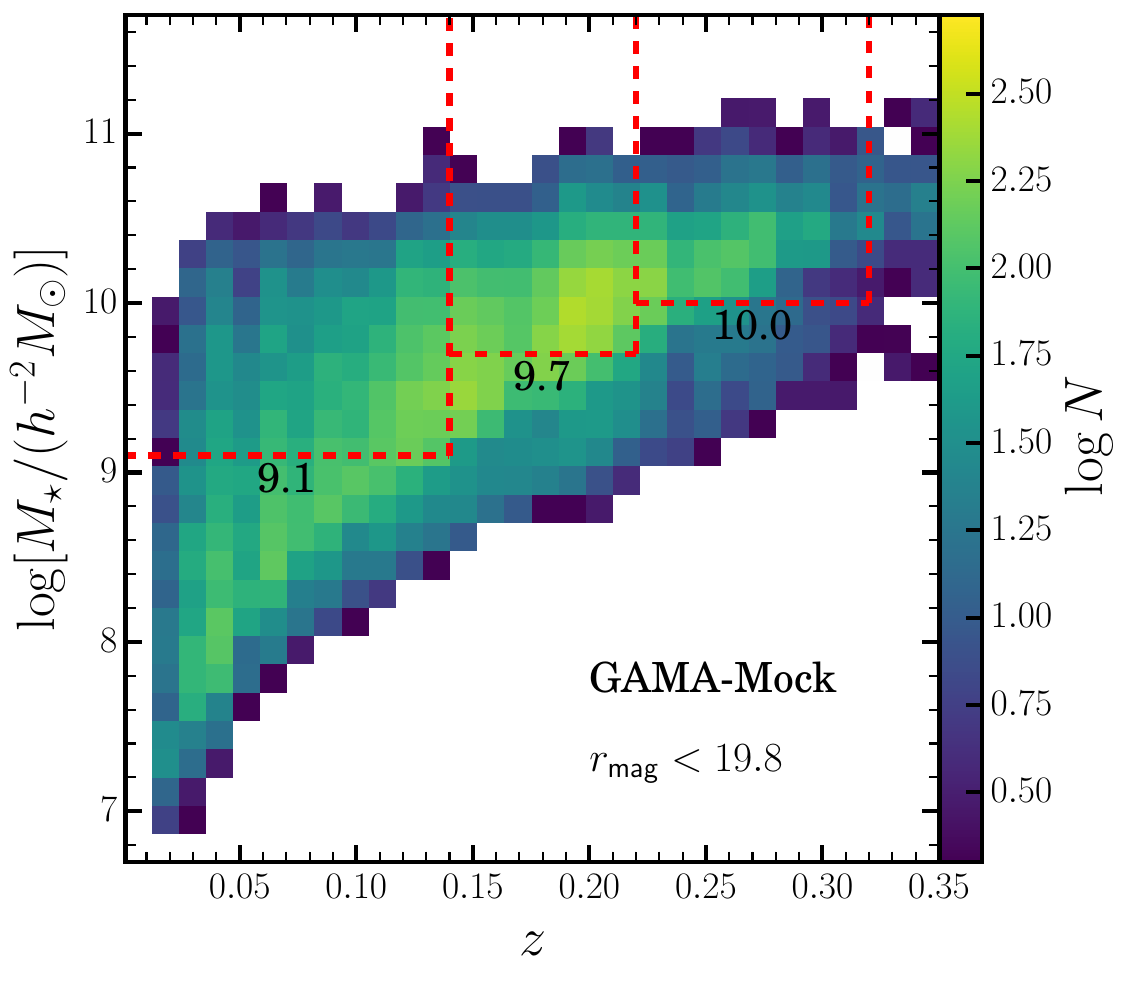}  
    \includegraphics[page=2, width=0.95\columnwidth]{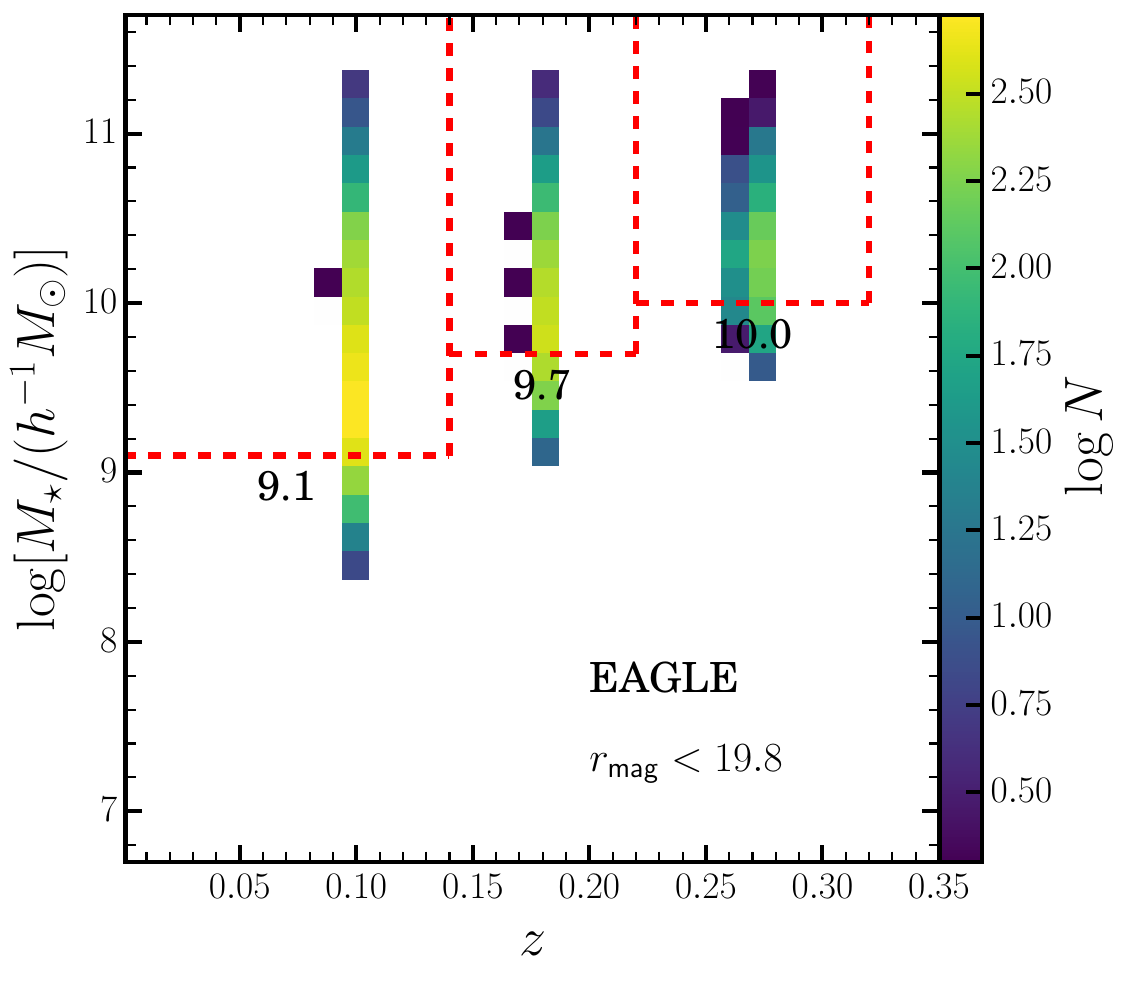}   
   \caption{Determination of the stellar mass completeness as a function of redshift.
            The panel at the top shows \gama\ data, the middle-panel shows the \gamamock\ and the bottom one shows \eagle\ data.
            All panels show stellar mass-redshift joint distributions colour coded by counts in the log scale. 
            Dashed red lines show stellar mass complete sample in different redshift windows. 
            The white dashed line on the top-left panel is the running $90^{\text{th}}$ percentile of the distribution.}
\label{fig:stellarmasscut}
\end{figure}
The shortcoming of a magnitude limited survey like \gama\ is that it observes a small (large) volume for the less (more) luminous galaxies.
In other words, the mass completeness limit of the survey varies as a function of redshift.
Figure~\ref{fig:stellarmasscut} shows the stellar mass-redshift relation for all data sets
highlighting the varying mass incompleteness as a function of redshift.
To tackle the bias introduced by this incompleteness (Malmquist bias), 
we adopt a conservative but robust approach of sub-selecting a volume complete sample. 
For this we have to first estimate a reasonable lower limit on stellar mass as a function of redshift. 
This is determined using the running $90^{\text{th}}$ percentile of the stellar mass distribution of the galaxies in groups at all redshifts.
For the discussion on the choice of percentile and its effect on our final results refer to the Appendix~\ref{sec:effectofvollimit}.
In summary the precise choice of stellar mass limit has no discernible impact on our primary results concerning stellar mass segregation.

We show the stellar mass completeness boundary with the white line in the top panel of Figure~\ref{fig:stellarmasscut}. 
The coloured pixels in the figure show the joint distributions of the galaxy stellar mass and redshift, 
where the colour scale depicts the logarithmic number count of galaxies in the underlying pixel. 
The redshift ranges we use are $0 < z \leqslant 0.14$, $0.14 < z \leqslant 0.22$ and $0.22 < z \leqslant 0.32$, 
and are chosen such that their mid-values are roughly equal to the redshift corresponding to the available snapshots of the \eagle\ simulation. 
In these redshift ranges for the \gama\ data, we determine the minimum 
complete log stellar mass values to be of 9.1, 9.7 and 10.0 respectively. 
The horizontal red lines in the figure are the demarcation of the lower-bound in the stellar mass at each redshift range. 
We will also present results for a single $z \leqslant 0.32$ range, where we assume a conservative mass completeness lower-limit of $\log(M_\star/M_\odot) = 10.0$.   

The synthetic \gamamock\ and \eagle\ data are complete down to the resolution limit of the simulation. 
However, for an effective comparison with the observed data, we impose a magnitude limit of $r_\text{mag}<19.8$ mag
(identical to the observed \gama\ data).
To calculate stellar mass limits for the simulated data (\gamamock\ and \eagle)
we repeat the same exercise as in the \gama\ data.
In the redshift ranges given above even for the simulated data, 
we find the lower limit in stellar mass to be similar to the \gama\ data. 
However, to make the final analysis comparable we impose exactly the same minimum limit on the stellar mass 
of the \gama\ data to all the three sets of data. 
The middle and bottom panels of Figure~\ref{fig:stellarmasscut} shows the 
the stellar mass-redshift joint distributions for our \gamamock\ and \eagle\ samples.
Again, the horizontal red lines show the demarcation of the lower limit in the stellar mass at each redshift range
whereas the vertical red lines divide the redshift range in which latter we study mass segregation. 
Note, the discreteness in redshift seen in the bottom panel, which shows the \eagle\ data, 
is due to the fact that the hydrodynamical simulation provides  
snapshots of the simulated universe at the discrete redshifts.

\section{Results}\label{sec:result}

\begin{figure*} 
   \centering
    \includegraphics[width=2.15\columnwidth]{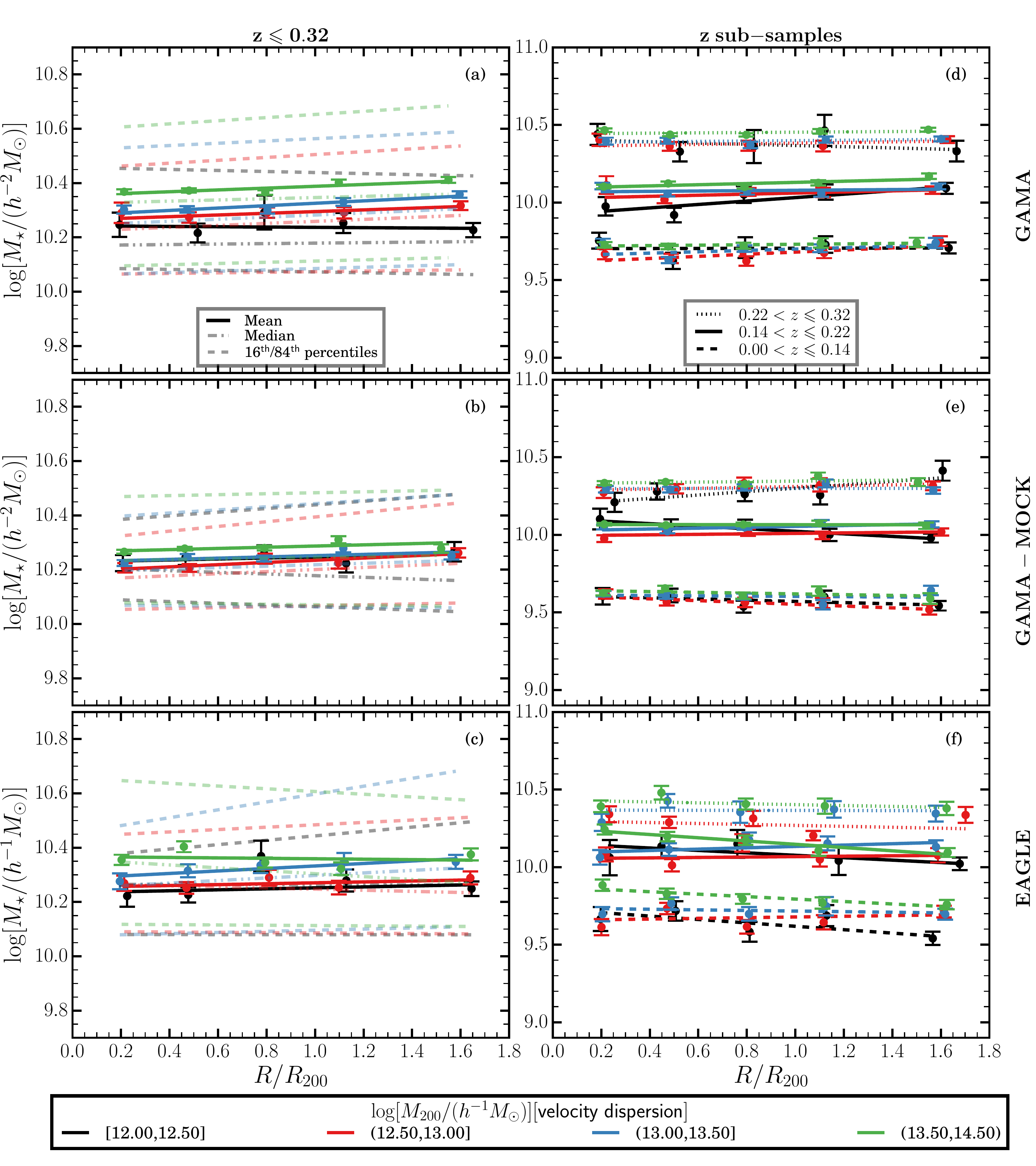}
    \caption{Stellar mass distributions of satellite galaxies in the galaxy group catalogues taken from \gama\ (top row), 
            \gamamock\ (middle row) and \eagle\ (bottom row). 
            In all panels, different colours represent different halo mass range.
            The left column shows the radial runs of the central-tendency of $\log(M_\star)$ of the galaxies 
            in galaxy groups of different halo mass ranges for $z\in(0.00,0.32]$ with $\log(M_\star) \geqslant 10.0$.
            Means of the $\log(M_\star)$ are shown with solid lines and medians are shown with faint   
            dashed-dotted lines whereas the faint dashed lines above(below) the solid lines are the
            $84^{\text{th}}$($16^{\text{th}}$) percentile of the $\log(M_\star)$ in a given data range.
            The column in the right side shows expectations of $\log(M_\star)$ in different redshift sub-samples as a function of halo mass.
            Here, the dashed, solid and dotted lines represent means of the $\log(M_\star)$ in 
            $z \in (0.00,0.14] \text{ with } \log(M_\star) \geqslant 9.1$, $z \in (0.14,0.22] \text{ with } \log(M_\star) \geqslant 9.7$ 
            and $z \in (0.22,0.32]  \text{ with }  \log(M_\star) \geqslant 10.0$ respectively.}
\label{fig:massseg_all}
\end{figure*} 
The main aim of this study is to investigate mass segregation of satellite galaxies 
in galaxy groups, and its dependence on halo mass.
An additional by-product of this is the investigation of the redshift evolution of the spatial distribution of galaxy mass in groups.
In general, there are very few high multiplicity groups within the observational limits of the \gama\ ($r_{\text{petro}}<19.8$),
and likewise in the \gamamock\ and \eagle\ data once the magnitude limit is applied,
which means we are unable to study stellar mass distributions on a group by group basis. 
To enhance the signal we stack groups and study their average properties instead,
as done in previous similar studies \citep[e.g.][etc]{2012MNRAS.424..232W,2013MNRAS.434.3089Z,2015MNRAS.448L...1R}.
For stacking, we scale the group-centric projected distance $R$ of the galaxy by the virial radius $\rvir$ of the group,
which should make the group size scale free.
We then investigate the stacked properties of group galaxies in halo mass 
$\logmhalo \in [12.0, 12.5],(12.5,13.0], (13.0, 13.5]$ and $(13.5, 14.5)$ ranges.

\subsection{Analysis of mass segregation in galaxy groups}
First, we study how the stellar mass, $\log(M_\star)$, of the satellite galaxies in groups varies with 
the scaled group-centric radii $R/\rvir$ in each of our halo mass $\logmhalo$ ranges.  
In the left column of Figure~\ref{fig:massseg_all}, we show the mean (by solid lines) and 
median (by faint uneven dashed lines) values of the $\log(M_\star)$ of satellite galaxies in different $R/\rvir$ and $\mvir$ ranges.
The top (panels a and d), middle (panels b and e) and bottom (panels c and f) rows demonstrate the distributions of the \gama, \gamamock\, and \eagle\ data respectively.
The $\logmhalo \in [12.0, 12.5],(12.5,13.0], (13.0, 13.5]$ and $(13.5, 14.5)$ are shown in black, red, blue and green colours respectively. 
Furthermore, the faint dashed lines below and above the solid dashed lines show the $16^{\text{th}}$ and $84^{\text{th}}$ percentiles of the distribution.
The redshift range of the data in this case is $z \leqslant 0.32$ and limited to $\log(M_\star) \geqslant10.0$ to guarantee stellar 
mass completeness (as described in Section~\ref{sec:vollimit}).
Moreover, we divide the data in five $R/\rvir$ ranges: $(0.0, 0.32], (0.32, 0.64], (0.64, 0.96], (0.96, 1.28]$ and $(1.28, 2.0)$.
Error bars shown in the data points in all the figures are the standard error of means for samples in the given halo mass and scaled radius ranges. 
For all of our mass segregation trends we fit a straight line with the uncertainties taken into account as described in \cite{2010arXiv1008.4686H,2015PASA...32...33R}.
The solid, dashed or dotted lines shown in all the figures throughout the paper show the resultant best fit models. 

There are few common trends that emerge from all three sets of data presented in panels (a), (b) and (c) of Figure~\ref{fig:massseg_all}. 
First, both the mean (solid lines) and median values (faint uneven dashed lines) of $\log(M_\star)$ show consistently similar trends. 
Thus, for clarity, and also for a sake of convenience in comparing with the literatures, 
we subsequently highlight the mean $\langle\logmhalo\rangle$ and show other central tendencies such as medians and percentiles with fainter lines.
Secondly, the mean trend lines (solid lines) show negligible gradients with the scaled-radius out to twice the group virial radii.
As such we fail to detect mass segregation ubiquitously for all the three data sets in the redshift range $z \leqslant 0.32$. 
Moreover, the absence of mass segregation trends seems independent of the halo mass range.
The slopes of our best fit mass segregation trends for all three data sets are $\lesssim 0.04$ dex.

We note in all three panels (a), (b) and (c) of Figure~\ref{fig:massseg_all} that the $16^{\text{th}}$ percentile of the $\log(M_\star)$ distributions 
(shown by faint dashed lines sitting below the solid lines), 
irrespective of the halo mass range, are clumped together near the limiting mass of our volume limited sample.
This is due to a hard lower-limit set on the stellar mass as a function of redshift (Figure~\ref{fig:stellarmasscut}).
On the other hand, the $84^{\text{th}}$ percentile of the $\log(M_\star)$ distributions 
(faint dashed lines sitting above the solid lines in the figure) show higher normalisation with increasing halo mass. 
For example, for the \gama\ data (panel a) the green lines representing the highest $\logmhalo$ group are above the blue lines, 
followed by red with the black line representing the smallest $\logmhalo$ groups at the bottom.
It is due to a complex combination of occupation physics that we observe higher mass halos hosting, on average, more massive galaxies.
This is a well understood effect \citep[e.g.][etc]{2006ApJ...647..201C,2010ApJ...717..379B,2010ApJ...710..903M,2010MNRAS.404.1111G,2013ApJ...770...57B}
and is also observed in the earlier Figure~\ref{fig:mstar_halo}, 
where the $\log(M_\star)$ versus $\logmhalo$ joint distributions have positive gradients.

\subsection{Lack of evidence of redshift evolution in satellite stellar mass distribution}
Here, we investigate the redshift evolution of the distribution of the stellar masses in galaxy groups. 
For this analysis, we separate our data, in particular, \gama\ and \gamamock, into the three redshift ranges 
$z \in (0,0.14], (0.14,0.22]$ and $(0.22,0.32]$.
As discussed in Section~\ref{sec:vollimit}, these ranges are chosen such that their mid-values roughly equal to the 
redshift corresponding to the available snapshots of the \eagle\ simulation.  
Also, as discussed earlier (again in Section~\ref{sec:vollimit}), to avoid the Malmquist bias
the above samples are then stellar mass limited to $\log(M_\star) \leqslant 9.1$, $\leqslant 9.7$ and $\leqslant 10.0$ respectively.

All panels (d), (e) and (f) in the right column of Figure~\ref{fig:massseg_all} 
show the redshift evolution of the stellar mass distribution in the groups out to twice the group virial radius.
The dashed, solid and dotted lines here show the mean $\log(M_\star)$ in increasing order of redshift.
As mentioned earlier, the mean and median values of $\log(M_\star)$ are consistent with each other and hence, 
here we only show mean values for clarity.
The different colours denote different halo mass ranges as labelled at the bottom of the figure.
Here we again we bin the data in radial ranges, 
$R/\rvir \in (0.00,0.32], (0.32,0.64], (0.64,0.96], (0.96,1.28]$ and $(1.28,2.0]$.   

In panels (d), (e) and (f) of Figure~\ref{fig:massseg_all}, a clear normalisation shift of $\log(M_\star)$ as a function of redshift can be seen. 
The shift is an artefact introduced due to the different lower limits on $\log(M_\star)$ imposed on the data for 
different redshift brackets while creating a stellar mass-limited sample
(red-dashed lines in Figure~\ref{fig:stellarmasscut}).
This systematically offsets the mean values, i.e. at high redshifts we do not detect the lower mass galaxies and hence the mean stellar mass is higher. 
There are a few ways to rectify this effect. 
For example, instead of $\log(M_\star)$ one could use $\log(M_\star)$ scaled by central satellite galaxy mass or 
the $\log(M_\star)$ renormalised by the median $\log(M_\star)$ values of the distribution of galaxies in each red boxes from the corresponding panels in Figure~\ref{fig:stellarmasscut}.
Moreover, one can also fit a stellar mass function \cite[e.g.,][etc]{2012MNRAS.421..621B, 2016MNRAS.457.1308M,2016arXiv160400008W} 
separately for all redshift ranges and then scale the $\log(M_\star)$ by the obtained break mass. 
However, since the main objective of our work is to investigate the gradient of the distributions and not their 
normalization, we leave the distributions unscaled and note this effect. 

In Figure~\ref{fig:massseg_all}(f) for \eagle\ data with $\log(M_{200}) \in(12.00,12.5]$ and $z\in(0.00,0.14]$, shown with black dashed line, 
we see a mild segregation trend with a gradient of $-0.11\pm{0.06}$ dex.
Also, we note that the \gamamock\ data (Figure~\ref{fig:massseg_all}e) 
with $z \in (0.22,0.32]$ at $R/\rvir<0.5$ and $\logmhalo \in [12.00,12.50]$  
show a strange increasing trend with slope $0.11\pm{0.05}$ dex.
This is in a contrast to what we observe in the corresponding \gama\ data in panel (d) and for \eagle\ data in panel (f). 
This contradicting behaviour is due to the difference in $\log(M_\star)$ and $\logmhalo$ joint distributions at $z\in(0.22,0.32]$ in the top and mid panels in Figure~\ref{fig:stellarmasscut}.
Overall, comparing panels (d), (e) and (f) of Figure~\ref{fig:massseg_all} we conclude that in overall there is negligible mass segregation in the 
groups with absolute gradient $\lesssim 0.08$ and consistent to zero when uncertainties in the slope is considered.
Interestingly, the satellite stellar masses as a function of scaled group radii for all the three data sets do not show any redshift evolution either. 

\subsection{Mass segregation in EAGLE data out to $r<22$ mag?}
\begin{figure*}
   \centering
      \includegraphics[width=2\columnwidth]{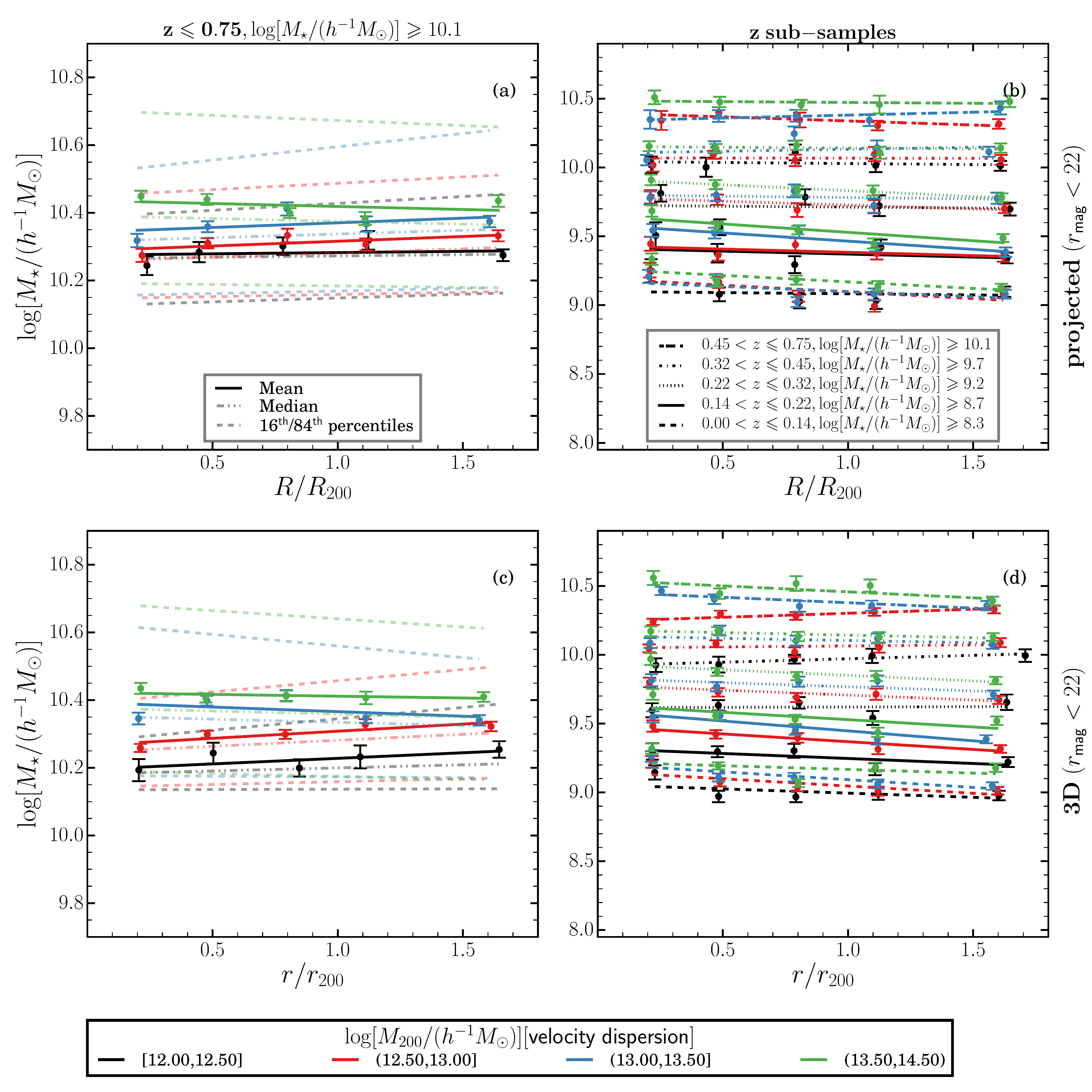}   
   \caption{Radial distribution of the stellar mass of the satellite galaxies in \eagle\ galaxy groups out to a fainter magnitude limit of $r_{\text{mag}}<22$. 
            Top panels show distributions in projected space, i.e, in observational space with inferred values for the masses.
            Bottom panels show the distributions using the full 3D information of the galaxies in groups and also, using the intrinsic values for stellar and halo masses.
            The meanings of the different line types in the above figure are identical to Figure~\ref{fig:massseg_all}.}
   \label{fig:eaglemagextra}
\end{figure*} 
The \eagle\ data can give us more insights into the stellar mass distribution of satellite galaxies in groups beyond 
what current observable data can offer. 
In particular, it allows us to probe galaxies, and hence groups, at fainter magnitude, 
and to observe stellar mass distributions in 3D space
with theoretically intrinsic values for key quantities such as $\log(M_\star)$ and $\logmhalo$ instead of the 
estimated values based on simple observed scaling relations. 

The stellar mass resolution limit of the \eagle\ simulation is $\log(M_\star) > 8.2$.
This means we can probe to a fainter magnitude limit of $r_{\text{mag}}<22$ allowing us to make predictions that can be tested with group catalogues generated 
from the future redshift surveys such as WAVES \citep{2015arXiv150700676D}.  
The apparent magnitude limit to $r<22$ mag means
we can now study satellite mass distribution out to a redshift $z \leqslant 0.75$.
Results shown in the top panels (a) and (b) of Figure~\ref{fig:eaglemagextra} are obtained repeating the same analysis as 
in the bottom panels (c) and (d) respectively of Figure~\ref{fig:massseg_all}, but with the fainter magnitude limited sample of $r<22$ mag.

In addition, \eagle\ also provides the full 3D distributions of the galaxies in groups.
In Figure~\ref{fig:eaglemagextra} (c) and (d), we show the mass segregation in \eagle\ groups
for $r_{\text{mag}}<22$ using the intrinsic values for the $\log(M_\star)$, $\logmhalo$, and 
spherical radius ($r$) instead of the projected radius.
We undertake this exercise with an ideal data set to highlight that the lack of mass segregation is possibly physical and not simply the manifestation of projected data and 
the approximate galaxy/group properties we use in reality.  
Panels (a) and (c) of the figure show spatial distributions of the satellite mass $\log(M_\star)\geqslant10.1$ in 
the given ranges of host halo mass to a redshift range of $z\leqslant0.75$.
In the figure, we once again show the mean, median and the percentiles ($16^{\text{th}}$ at the bottom, $84^{\text{th}}$ at the top) of the $\log(M_\star)$
by solid, faint dashed-dotted and faint dashed lines respectively.
Similarly, panels (b) and (d) show the spatial distributions of the satellite mass 
$\log(M_\star)$ in the given ranges of host halo mass in 5 different redshift ranges.
For clarity we do not show the percentiles in the right side panels. 
In all panels, different colours represent different host halo mass ranges. 

Overall, in panels (a) and (b) we again observe negligible mass segregation with absolute gradient of $\lesssim0.03$ dex 
and $\lesssim 0.09$ dex respectively. 
Similarly, in panel (c) and (d), which uses the ideal data, we still do not observe any radial gradients in stellar mass runs.
All these results hold irrespective of the host halo mass.
However, in panels (b) and (d) for the cases $z \leqslant 0.22$ there is seemingly some mild mass segregation 
with slopes ranging between $0.06-0.1$ dex albeit 
with large uncertainties of typically $40\%$. 
Given that at $z \leqslant 0.22$ we are closer to the mass resolution of the \eagle\ simulation \citep{2015MNRAS.446..521S}  
it is difficult to be certain that this result is robust to simulation resolution limits.
However, in all other cases, i.e., $0.75\geqslant z>0.22$ the stellar mass distribution is almost flat once again demonstrating the lack of mass segregation. 

As evidenced in Figure~\ref{fig:halo_200}, there are uncertainties on the estimated virial radius $R_{200}$.
Hence, it is possible that for a large enough uncertainty on $R_{200}$, and therefore also in scaled radius $R/R_{200}$, any real underlying radial trend could be erased.
Due to a lack of intrinsic/true measurements of $R/R_{200}$ it is difficult to simulate this effect on \gama\ and \gamamock.
The \eagle\ data would be useful here as both the intrinsic and projected/noisy $R/R_{200}$ information are available. 
However, in the case of the \eagle\ data we have already seen in Figure~\ref{fig:eaglemagextra} 
that there is no segregation trend even when the intrinsic properties are consider.
Therefore, for this simulation we generate a synthetic stellar mass for \eagle\ galaxies sampled from a 
straight line of gradient $=-0.3$ dex, roughly of the same magnitude as seen in some literature, 
and also, introduce a normally distributed scatter of 0.35 dex around the line, which is a function of intrinsic $R/R_{200}$.
We then fit a straight line to the synthetic stellar mass as a function of noisy/observed $R/R_{200}$ with inherent error distributions as shown in Figure~\ref{fig:halo_200}.
Given the uncertainties in $R/R_{200}$ we were still able to recover the slope with $\lesssim10\%$ uncertainty. 
The uncertainty is close to $10\%$ for the lowest halo mass range whereas slightly smaller for the highest halo mass range. 
The above exercise suggests that the associated uncertainties in the derived virial properties 
do not erase the signal unless the gradient is as tiny as 0.03 dex.

\section{Discussion}\label{sec:discussion}

\subsection{Robustness of the absence of mass segregation to different halo mass estimates}
As discussed earlier, the total dynamical mass of a group can be estimated in numerous ways, such as
from its velocity dispersion using the virial theorem \cite[e.g.][also Equation \ref{eqn:m200}, etc]{1997ApJ...485L..13C,2006eac..book.....S},
from weak gravitational lensing \cite[e.g.][etc]{1996ApJ...466..623B,2005ApJ...634..806P,2015MNRAS.452.3529V},
from its luminosity assuming some light-to-mass ratio
or from abundance matching \citep[e.g.][etc]{2010ApJ...717..379B,2010ApJ...710..903M,2013ApJ...770...57B,2013MNRAS.433..659H} etc.
An independent measurement of the halo mass using different methods for all three data sets
is a massive undertaking and well beyond the scope of this work.
However, the \gama\ group catalogue readily provides some alternative measurements of halo mass. 
Thus, here we confine our study to only the \gama\ group catalogue.

\begin{figure*} 
   \centering
   \includegraphics[width=2.1\columnwidth]{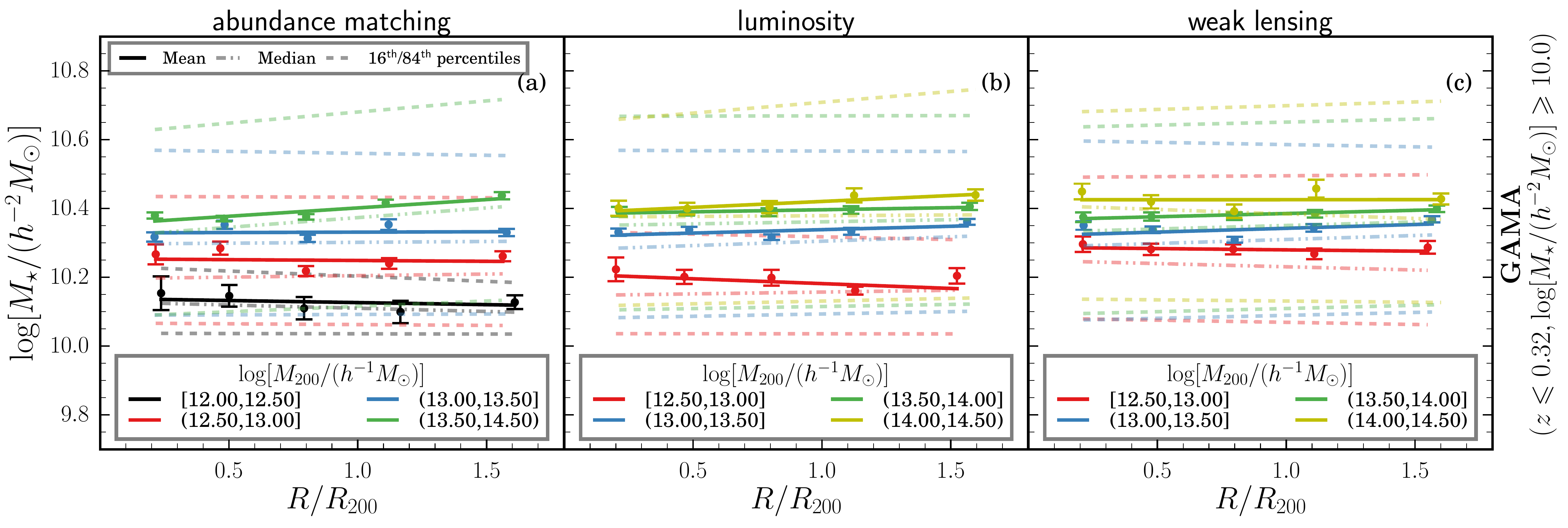}  
   \caption{Stellar mass distribution of the satellite galaxies in \gama\ galaxy groups 
            in different halo mass ranges for three halo mass definitions 
            (a) abundance matching, (b) luminosity based and (c) weak lensing.
            All labellings are identical to Figure~\ref{fig:massseg_all}(a),
            but note largest halo mass bin further split into two bins (13.50,14.00] and 
            (14.00,14.50).}
\label{fig:massseg_halosensitivity}
\end{figure*}
For reference we remind the reader that the mass segregation trends in \gama\ groups 
with virial theorem (velocity dispersion) based halo masses are presented in Figure~\ref{fig:massseg_all}(a).
Now, in Figure~\ref{fig:massseg_halosensitivity} we show the same as 
Figure~\ref{fig:massseg_all} but with halo mass $\logmhalo$ measured from three additional methods, (i)
 abundance matching using halo occupation distribution \citep{2012ApJ...745...16T}, 
(ii) group total luminosity \citep{2011MNRAS.416.2640R} and 
(iii) the weak lensing calibrated relation\citep{2015MNRAS.452.3529V}. 
Results from these different halo mass estimates are shown in panels (a), (b) and (c) of Figure~\ref{fig:massseg_halosensitivity} respectively.
Note, the $R_{200}$ used here are also recomputed for the different definition of $\logmhalo$ using Equation~\ref{eqn:m200}.
In panels (b) and (c), due to sparse data we are only able to probe above $\logmhalo \geq 12.5$, whereas
due to an increase in number counts in $\logmhalo \in (13.00,14.50]$ we split it into further two bins (13.00,13.50] and (13.50,14.50).
For all three cases typical values of the slopes of the mass segregation is $0.02\pm{0.02}$ dex meaning
there is still no segregation in \gama\ data highlighting that the lack of observed mass segregation is not due to our choice of halo mass estimator.
The exception to this is the case of $\logmhalo \in [12.5,13.0]$ range 
(Figure~\ref{fig:massseg_halosensitivity}b) and $\logmhalo \in [13.5,14.5]$
(Figure~\ref{fig:massseg_halosensitivity}a)
where we do see some mild segregation trend 
with slope of $\sim|0.04|$ dex, which we did not detect using the dynamically implied halo masses.
It could potentially be due to sample size fluctuation as a result of scatter between halo mass estimates obtained using different methods.
Overall, the trends for larger halo masses for all four halo mass measurements are broadly consistent.
This comparative study gives us confidence that the lack of mass segregation, at least in the case of \gama\ observational data, is robust to the halo masses used.

\subsection{Robustness of the absence of mass segregation to different group centre definitions}\label{sec:cendef}
\begin{figure} 
   \centering
   \includegraphics[width=0.98\columnwidth]{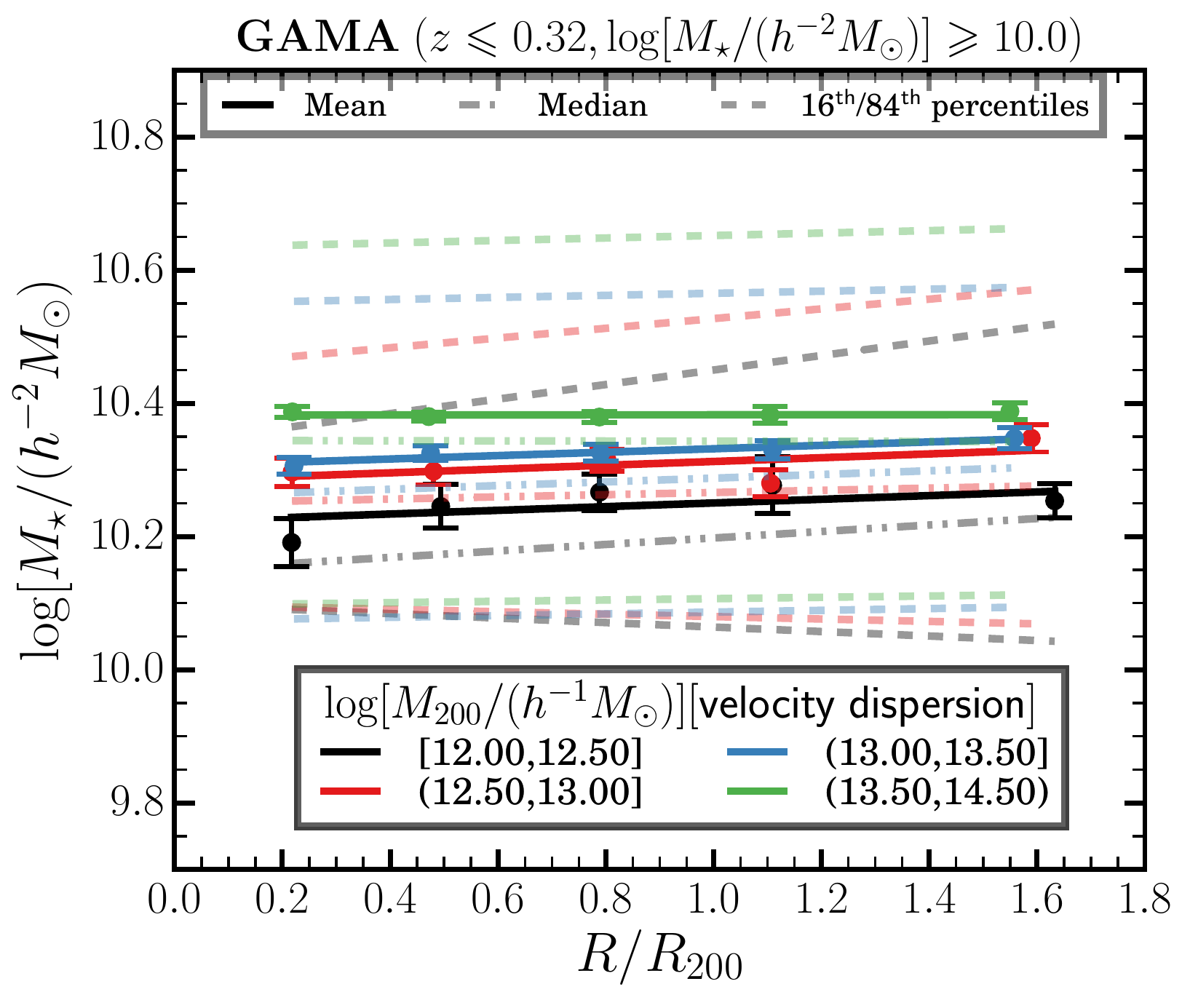}  
   \caption{Stellar mass distribution of the satellite galaxies in \gama\ galaxy groups with luminosity weighted centre. 
            All labellings are identical to Figure~\ref{fig:massseg_all}(a).}
\label{fig:cendef}
\end{figure}

As discussed earlier, a centre of any galaxy group can be pinned 
to be at its luminosity weighted centre or at the location of its brightest group galaxy (BGG).
The \gama\ group catalogue provides both the measurements of group centre.
Here, we use them to test the robustness of the absence of mass segregation to the different definitions for the group centre. 
Figure~\ref{fig:massseg_all}(a) already shows the mass segregation in \gama\ groups assuming BGGs as the group centres. 
Therefore in Figure~\ref{fig:cendef} we repeat the analysis for \gama\ groups 
assuming the luminosity weighted centre (labelled as Cen in the catalogue). 
Except for the lowest halo mass bin where scatter in the data is large, slopes of the trends for all the other halo 
mass ranges are consistent with zero. 
Comparing Figures~\ref{fig:massseg_all}(a) and \ref{fig:cendef} it can be concluded that 
the effect of the above mentioned choices of the group centre is on average
negligible on the mass segregation trends in \gama\ data.

\subsection{Mass segregation in SDSS?}\label{sec:mass_seg_sdss}
Here we compare our results to mass segregation studies reported in the literature that use SDSS group catalogues.
In a study using a $V_\text{max}$ weighted sample based on the SDSS DR4 galaxy group catalogue of Y07, \cite{2008arXiv0805.0002V}
find a significant mass segregation gradient of $\sim0.5$ dex over an extent of one virial radius 
\footnote{It should be noted that the range of the scaled radius varies
depending on the choice of over-density constant $\Delta$. 
For example, the distance where the mean matter density of the group is 
$\Delta=200$ times the mean background matter density will always be smaller than when $\Delta=180$ is assumed.
It means for the same range in $R$, $R/R_{180}$ spans to smaller range compare to $R/R_{200}$.}.
Additionally, they find that the mass segregation occurs at all halo mass ranges. 
Interestingly, more recent work in \cite{2015MNRAS.448L...1R}, again based on Y07 (but using SDSS DR7), 
presents a slightly different picture than the previous work.
For example, for their case of $\log(M_\star) > 9.0 + V_\text{max}$, which is equivalent to \cite{2008arXiv0805.0002V} studies, 
we see that the magnitude of spatial mass segregation is $\lesssim0.2$ dex for the
low halo mass ($\logmh<13$) case, and it is almost negligible ($\lesssim0.05$ dex) for larger halos. 
In contrast to above works, using galaxy group catalogue created from SDSS DR7 but with modified implementation of the Y07 group-finder, 
\cite{2012MNRAS.424..232W} fail to detect evidence for satellite mass segregation at any halo mass range.
This result is in overall agreement to our findings of absence of mass segregation in \gama, \gamamock\ and \eagle\ galaxy groups.

There could be various potential reasons resulting in the contrasting mass segregation trends. 
For example, differences in arbitrary stellar mass completeness limit in previous studies, 
the subtleties of group-finding algorithms,
different prescriptions for stellar/halo masses being used, 
different definitions for the group centres or, potentially the combination of all the above possibilities. 
For example, the mass segregation trends with the conservative stellar mass limited (e.g., $\log(M_\star)>10.5$) case
compared to the volume-weighted $\log(M_\star) >9.0 + V_\text{max}$ case in Figure 1 (note different range of y-axis in two panels) of \cite{2015MNRAS.448L...1R}
are much steeper than the latter case.

Similarly, \cite{2012MNRAS.424..232W} and \cite{2015MNRAS.448L...1R}, who use the same input data, i.e., SDSS DR7, 
still find contradicting mass segregation trends.
The main difference between the two works is in their implementations of the Y07 group finder.
While the later work uses the original group catalogue, the earlier work uses a modified version of Y07. 
This means the subtle difference in the implementation of the group finding algorithms could be a factor.

In the following, we further investigate the source of contradicting results existing in the literature.
For this, first and foremost we adopt an independent group catalogue of SDSS data by \cite{2015arXiv151105856S},
constructed using the group finding algorithm similar to \citet[the \gama\ group catalogue]{2011MNRAS.416.2640R}.
Both studies use the friend-of-friends algorithm
with similar values for the linking lengths ($b_\bot \simeq0.06$ and 
$b_{\|} \approx1.0$), which are the distances that define which objects 
should be linked into common halos/groups.
These linking lengths are tuned to reproduce the properties of mock groups. 
Importantly, the linking lengths used in both the above works are also the values  
recommended from the recent investigation on the performance of friends-of-friends algorithm 
among various existing group catalogues by \cite{2014MNRAS.440.1763D}. 
\begin{figure*}
   \centering  
   \includegraphics[width=2.1\columnwidth]{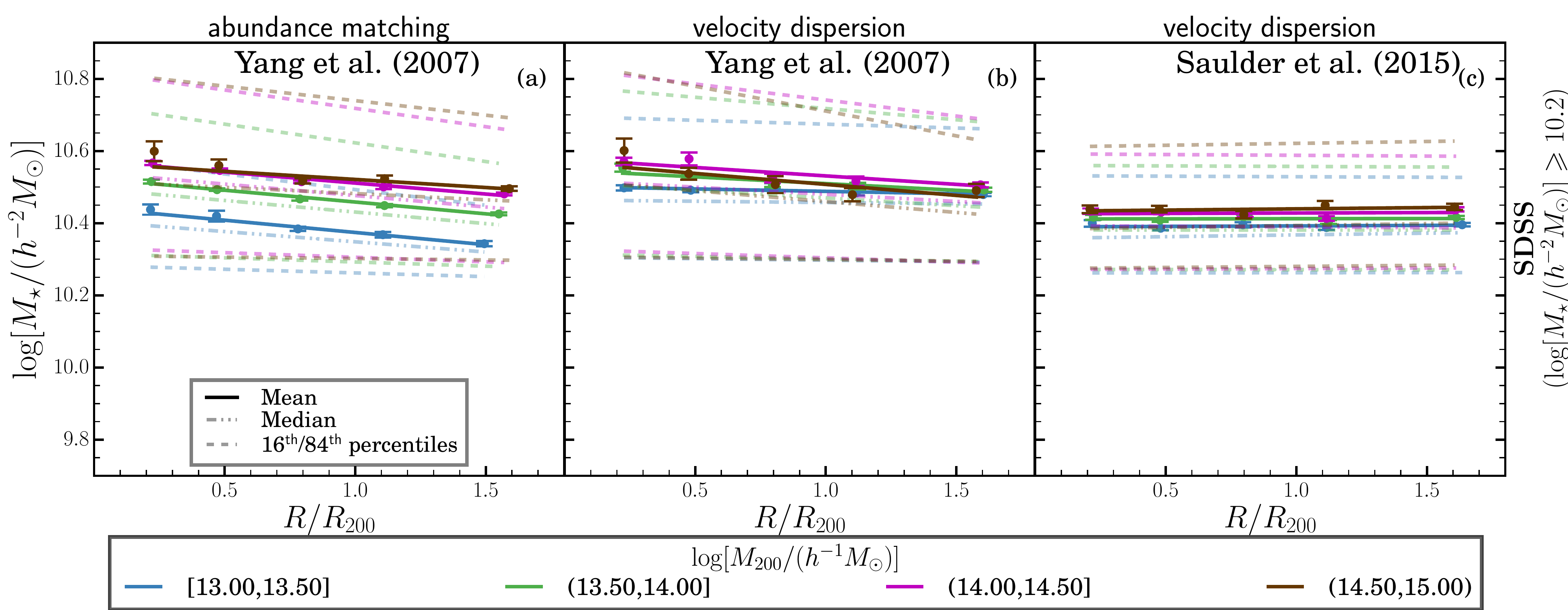}   
   \caption{Spatial mass distribution in SDSS galaxy groups as a function of host halo mass.
	    Panel (a) and (b) show results with SDSS group catalogue of 
	    \protect\cite{2007ApJ...671..153Y} using abundance matched and velocity dispersion based halo masses respectively.
	    Panel (c) shows results with SDSS group catalogue of \protect\cite{2015arXiv151105856S} using velocity dispersion based halo masses.
            The samples are stellar mass limited to $\log[M_\star/(h^{-1} M_\odot)]\geqslant10.2$.
            Mean values are shown with bold solid lines whereas faint lines represent percentiles as in Figure~\ref{fig:massseg_all}(a).
            Different colours show different halo mass ranges as depicted in the box at the bottom of the figure.}
\label{fig:massseg_sdss}
\end{figure*}
Secondly, to be consistent with our earlier observations with \gama, \gamamock\ and \eagle, 
we re-estimate physical properties such as galaxy and group masses, group radius etc 
for the \cite{2015arXiv151105856S} group catalogue with methods described in the Section~\ref{sec:derivedquant}.

In Figure~\ref{fig:massseg_sdss}(c) we present our results for SDSS data using the \cite{2015arXiv151105856S} group catalogue. 
Here, to facilitate comparison with previous works with SDSS \citep[e.g.][etc]{2015MNRAS.448L...1R}, we only investigate stellar mass limited sample of $\logmstar\geqslant10.2$.
Also note that, like in the above subsections here also we scale group radii with corresponding $\rvir$.
Since we are mainly interested in the gradient of the radial distribution of the satellite stellar masses
the effect of the choice of $\Delta=180$ versus 200 is minimal in the direct comparison between results in 
Figures~\ref{fig:massseg_all}(a), (b), (c) and \ref{fig:massseg_sdss}, and e.g. \cite{2008arXiv0805.0002V}.
In Figure~\ref{fig:massseg_sdss}(c) we demonstrate that the recent SDSS group catalogue of \cite{2015arXiv151105856S} 
also does not show any mass segregation, where the maximum value of the 
slope of trend lines for all the halo mass ranges is $0.01\pm0.01$. 
This is in an agreement with our findings from \gama, \gamamock\ and \eagle\ data and also, from the studies of SDSS data by \cite{2012MNRAS.424..232W}
whereas in contradiction with the other studies of SDSS data by \cite{2008arXiv0805.0002V} and \cite{2015MNRAS.448L...1R}.

The differences in the mass segregation trends observed in SDSS data by  
\cite{2012MNRAS.424..232W,2015arXiv151105856S} and by \cite{2008arXiv0805.0002V,2015MNRAS.448L...1R} 
could perhaps be due to the inaccuracies associated with the group finding in the very first place.  
To investigate this, here, we directly adopt the group catalogue of \citet[Y07]{2007ApJ...671..153Y} for SDSS data. 
In Figure~\ref{fig:massseg_sdss}(a) we demonstrate that we recover the mass segregation trend in the Y07 group catalogue.
The magnitude of mass segregation in Y07 here ranges from $-0.08\pm{0.01}$ dex for $\logmh\in[13.0,13.5]$ to $-0.05\pm{0.01}$ dex for $\logmh\in(14.5, 15.0)$ i.e.
segregation becomes sallower with increase in the halo masses, which are roughly consistent with earlier work by \cite{2015MNRAS.448L...1R}.
To produce this result, we take the galaxy and group properties from the published group catalogue of Y07. 
Note, the halo masses in the Y07 catalogue are based on two measurements: 
the total luminosity or total stellar mass of the all group members brighter than $M_r<-19.5$.
We find that using either of these two estimates for halo mass makes negligible difference in observed mass segregation.
In Figure~\ref{fig:massseg_sdss}(b) we repeat the same analysis 
with the group catalogue of Y07 but using implied dynamical halo mass measured from velocity dispersion as described in Section~\ref{sec:virialprop}, 
and consistent with the cases of \gama, \gamamock, \eagle\ and \cite{2015arXiv151105856S} SDSS group catalogue.
As a final check, for both the left and mid panels of Figure~\ref{fig:massseg_sdss}, 
we compute $\logmstar$ using the relations in Y07, which is a function of $g$ and $r$ band magnitudes instead of one given in Equation~\ref{eqn:stellarmass}.  
We find that the differences of these two colour based calibrations for galaxies stellar mass has negligible influence in our result. 
Comparing panels (a) and (b) of the figure, we note that the segregation is already reduced when switched to velocity dispersion based mass estimate.
In particular for a case of $\logmh<13.5$ in middle panel the segregation is almost negligible 
($\lesssim0.03$ dex from $\lesssim0.07$ dex).
This result is consistent with our findings from \gama, \gamamock\ and \eagle, strictly speaking to the cases of $z \leqslant 0.14$ given the shallower redshift range of SDSS data. 

If we compare Figure~\ref{fig:massseg_sdss}(b) with Figure~\ref{fig:massseg_sdss}(c)
we can see that there are still some segregation trends visible in the larger $\logmh>13.5$ cases. 
This could possibly be due to subtleties of group finding scheme,
linking lengths, or their complex combination.
Moreover, Y07 use an imprecise scheme to estimate the dependence of 
luminosity incompleteness on redshift in their flux-limited sample
as discussed in \cite{2015MNRAS.453.3848D} and is also apparent in Figure 4 of Y07. 
These errors in the luminosity incompleteness propagate to the inferred group masses.
Thus, the dramatic decrease in the segregation trend between Figure~\ref{fig:massseg_sdss}(a)
(equivalent to Figure 1 in \citealt{2008arXiv0805.0002V}) 
and Figure~\ref{fig:massseg_sdss}(b) appears to be due to the 
luminosity based halo mass measurements provided in Y07.

We note that \cite{2008arXiv0805.0002V,2015MNRAS.448L...1R} consider
the luminosity weighted centre as a group centre whereas \cite{2012MNRAS.424..232W},
Figure~\ref{fig:massseg_sdss}(c) using \cite{2015arXiv151105856S} 
and all of our analysis in the previous sections assume BGG as a group centre.
Similarly, the fact that in Figure~\ref{fig:massseg_sdss} (a) and (b) we are using BGG centres and still being
able to recover the segregation trend in \cite{2008arXiv0805.0002V, 2015MNRAS.448L...1R} suggest that 
the effect of the above definitions of group centres is negligible. 
This is also in agreement with our conclusion for \gama\ data presented in Section~\ref{sec:cendef}.

In summary, the segregation trends seen in the Y07 group catalogue of SDSS data shown 
in Figure~\ref{fig:massseg_sdss}(a) and also observed in earlier studies 
\citep{2015MNRAS.448L...1R,2008arXiv0805.0002V} are in contrast to our findings 
with \gama, \gamamock\ and \eagle\ group catalogues, 
as well as to the findings of \cite{2012MNRAS.424..232W} and our Figure~\ref{fig:massseg_sdss}(c) 
with the \cite{2015arXiv151105856S} \sdss\ group catalogue.
From the discussions in the above paragraphs, we deduce that perhaps the difference is inherently linked
to the construction of Y07 group catalogue.
As discussed in \cite{2015MNRAS.453.3848D} potentially the imprecise scheme of computing 
the luminosity incompleteness as function of redshift during the group finding in Y07, which eventually propagates to the abundance matching
technique leading to the incorrect estimate of group masses is a plausible culprit.
In the future it would be interesting to see how the improvement in the halo mass measurement for Y07 group catalogue
suggested in \cite{2015MNRAS.453.3848D} influences the spatial mass segregation results.

\subsection{Anti-segregation trend beyond the virial radius?}
In a recent study of semi-analytic models of galaxy formation, \cite{2015arXiv150308342C} report an interesting 
claim that beyond the virial radius there is a global and significant increasing trend in stellar mass.
They attribute this strange upturn to a presence of intrinsically massive and recently accreted objects at large radius. 
In our semi-analytic data \gamamock\ presented in Section~\ref{sec:result} (Figure~\ref{fig:massseg_all}) we note insignificant upturns (with gradient $\lesssim 0.02$ dex) in the satellite mass distribution 
beyond the virial radius.
This is in clear contrast to the findings of \cite{2015arXiv150308342C}.

\subsection{A comment on some observed spurious trends}
We find above that \eagle\ data with $z\in(0.00,0.14]$ and $\logmhalo\in(12.0,12.5]$
shown in Figure~\ref{fig:massseg_all}(f), \gamamock\ data with $z\in(0.22,0.32]$ and $\logmhalo\in(12.0,12.5]$ shown in Figure~\ref{fig:massseg_all}(e) etc
show strange trends in a contrast to our overall observation of lack of mass segregation. 
In order to put the apparent significance of any single measurement into context we analyse 
the distribution of all line fits normalised by the estimated error.
Here, we assume that all the measurements are independent, i.e., our fits are based on disjoint sub-samples.
\begin{figure}
   \centering  
   \includegraphics[width=0.9\columnwidth]{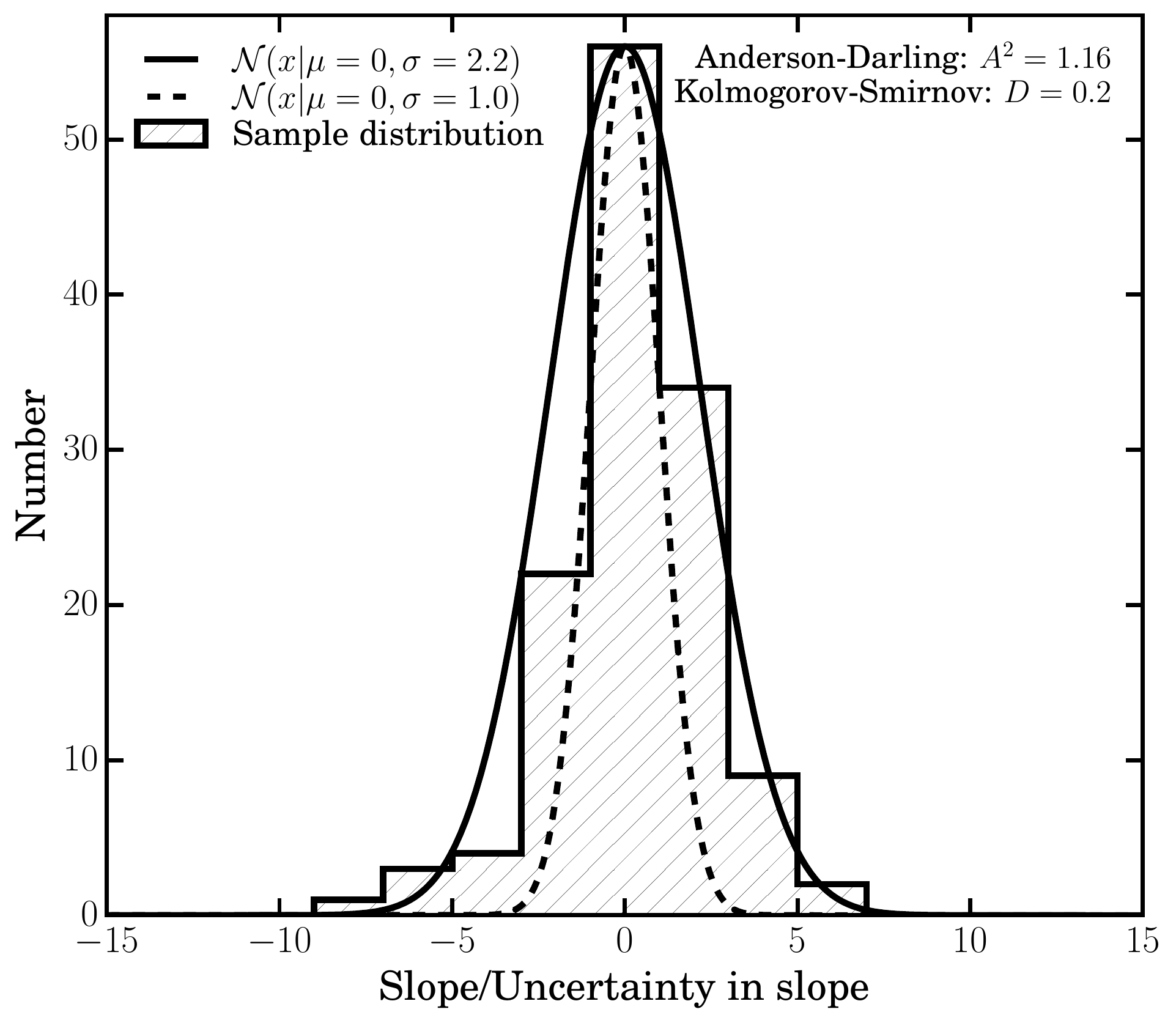}   
   \caption{Distribution of the slopes of the mass segregation trend lines normalised by the estimated error. 
            The solid and dashed lines are normal distributions with different values for dispersions overlaid for a reference.
            }
\label{fig:slope_error}
\end{figure}
This distribution can be seen in Figure~\ref{fig:slope_error}, which includes the slopes and their corresponding uncertainties
for all the trends ($N=132$) shown in Figures~\ref{fig:massseg_all}-\ref{fig:cendef} and Figure~\ref{fig:stellarmasscut}
but excluding panels (a) and (b) of Figure~\ref{fig:massseg_sdss}.

We except that such distribution is normal around the expected value. 
There are plethora of available statistical tests to identify departure of any distribution from normality -
the Anderson-Darling A$^2$ test, the Kolmogorov-Smirnov $D$ test, 
the Shapiro-Wilk $W$ test, the Lilliefors test to name a few.
Here, we only consider the first two tests. 
The data that significantly depart from a Gaussian distribution is expected \citep[cf. Section 4.7.4][]{book1158673}
to have Anderson-Darling A$^2$ value $>>1$ and also, the Kolmogorov-Smirnov $D$ $>>1/\sqrt{N}$.
For our distribution the values these tests result are 1.16 and 0.2 respectively, which both suggest
the distribution is consistent to being normal. 
As a consequence, any single ``significant'' result must be tempered by the large number of effective trials (i.e. distributions fit). 
Indeed, even the most extreme positive or negative trends are entirely consistent 
with being samples taken from this global distribution of slopes.
For a reference we simply overlay a standard normal probability distribution i.e. with mean $\mu=0$ and dispersion $\sigma=1$, 
shown with the dashed line in the figure. 
Similarly, the smooth solid line shows a normal distribution assuming mean and dispersion of the data.
Interestingly, in the figure we see that the sample distribution is broader than the standard gaussian distribution 
(shown with the dashed line). 
It means that the estimated uncertainties in the slopes in some cases could have been underestimated.

\section{Conclusion}\label{sec:conclusion}
We investigate the controversial issue of the presence, or lack thereof, of mass segregation in galaxy groups. 
We provide a comprehensive study of the radial distribution of stellar mass of the satellite galaxies in galaxy groups
for observations: the galaxy-redshift survey Galaxy and Mass Assembly 
(\gama); semi-analytics: the \gama\ lightcone mock catalogues (\gamamock) constructed using a model of galaxy formation by the GALFORM group,
and cosmological hydrodynamical simulation: the Evolution and Assembly of 
GaLaxies and their Environments (\eagle).

Overall, the absolute gradient of spatial mass segregation in galaxy groups is found to be insignificant ($\lesssim0.04$ dex). 
We find this to be consistent for all the three data sets at various halo mass ranges between 
$12 \leqslant \logmhalo <14.5$ and in the redshift range $0 \leqslant z \leqslant 0.32$. 
Analogous to the observed \gama\ data, we magnitude-limit both the synthetic data i.e. \gamamock\ and \eagle\ to $r<19.8$ mag, 
and carefully select stellar mass complete samples at given redshift intervals. 
We also find that the radial distributions of the stellar mass does not show any redshift evolution out to $z \leqslant 0.32$.
In cases where we separate data into different redshift ranges the absolute gradients of spatial mass segregation trends 
were slightly larger $\lesssim0.08$ dex but consistent to zero given the uncertainties in the slope.
Moreover, we find that our results at least for the \gama\ data are robust to different halo mass and group centre estimates. 

The \eagle\ data give us further insights by allowing us to probe fainter magnitude limit of $r_{\text{mag}}<22$ and also, 
to study the three-dimensional spatial distributions using the intrinsic stellar and virial masses. 
Except for the low redshift regime $z \leqslant 0.22$, even with the fainter magnitude limit of $r_{\text{mag}}<22$, we find that the 
\eagle\ data do not show any mass segregation in the halo mass range $12 \leqslant \logmhalo <14.5$ and out to $z \leqslant 0.75$. 
This remains the case for both the projected and intrinsic data alike. 

Intriguingly, the lack of mass segregation we observe is in contrast to what has recently been reported in 
\cite{2008arXiv0805.0002V, 2015MNRAS.448L...1R} with the SDSS group catalogues of \cite{2007ApJ...671..153Y}.
We find that the magnitude of mass segregation seen in earlier works with SDSS group catalogues 
reduces when we replace their original luminosity based halo masses with dynamically inferred masses. 
As advocated in \cite{2015MNRAS.453.3848D}, the original estimates for halo masses from 
abundance matching could have propagated uncertainties from how \cite{2007ApJ...671..153Y} group catalogues are constructed.
A subtle effect due to using halo based group finding instead of FoF based finding could also potentially result in observed mass segregation. 
Interestingly, our analysis based on the SDSS group catalogue of \cite{2015arXiv151105856S}, 
which uses a similar group-finder to \cite{2011MNRAS.416.2640R},
accompanied with implied dynamical halo masses, confirms the lack of 
significant evidence of mass segregation in low redshift galaxy groups. 
This is entirely consistent with our findings from \gama, \gamamock\ and \eagle\ group studies and with the conclusion of \cite{2012MNRAS.424..232W}
using a revised SDSS group catalogue.

The apparent lack of mass segregation in groups suggests that whatever processes might enhance the effect (e.g. dynamical friction, mergers etc)
is sub-dominant compared to competing and masking processes (e.g. long time-scales, star-formation, quenching etc).

\section*{ACKNOWLEDGEMENTS}
PRK is funded through Australian Research Council (ARC) grant DP140100395 and the University of 
Western Australia Research Collaboration Award PG12104401 and PG12105203.
PRK thanks Christoph Saulder for making the SDSS group catalogue available before the publication, and 
Sanjib Sharma and Weiguang Cui for discussions at earlier stages of this project. 
We also like to thank the referee for the comments that helped to improve the content of the paper.

\gama\ is a joint European-Australasian project based around a spectroscopic campaign using the Anglo-Australian Telescope. 
The \gama\ input catalogue is based on data taken from the Sloan Digital Sky Survey and the UKIRT Infrared Deep Sky Survey. 
Complementary imaging of the \gama\ regions is being obtained by a number of independent survey programs including 
GALEX MIS, VST KiDS, VISTA VIKING, WISE, Herschel-ATLAS, GMRT and ASKAP providing UV to radio coverage. 
\gama\ is funded by the STFC (UK), the ARC (Australia), the AAO, and the participating institutions. 
The \gama\ website is \url{http://www.gama-survey.org/}

We acknowledge the Virgo Consortium for making their simulation data available. 
The \eagle\ simulations were performed using the DiRAC-2 facility at
Durham, managed by the ICC, and the PRACE facility Curie based in France at TGCC, CEA, Bruyeres-le-Chatel.

To construct the \gamamock\ the DiRAC Data Centric system at Durham University, operated
by the Institute for Computational Cosmology on behalf of the STFC DiRAC HPC Facility
(\url{www.dirac.ac.uk}), was used. This equipment was funded by BIS National E-infrastructure capital
grant ST/K00042X/1, STFC capital grant ST/H008519/1, and STFC DiRAC Operations grant
ST/K003267/1 and Durham University. DiRAC is part of the National E-Infrastructure. The development
of the \gamamock\ was supported by a European Research Council Starting grant (DEGAS-259586)
and the Royal Society.

\bibliographystyle{mnras}
\bibliography{ms.bib}

\appendix
\section{Effect of stellar mass completeness limits}\label{sec:effectofvollimit}

\begin{figure*}
   \centering  
   \includegraphics[width=2.2\columnwidth]{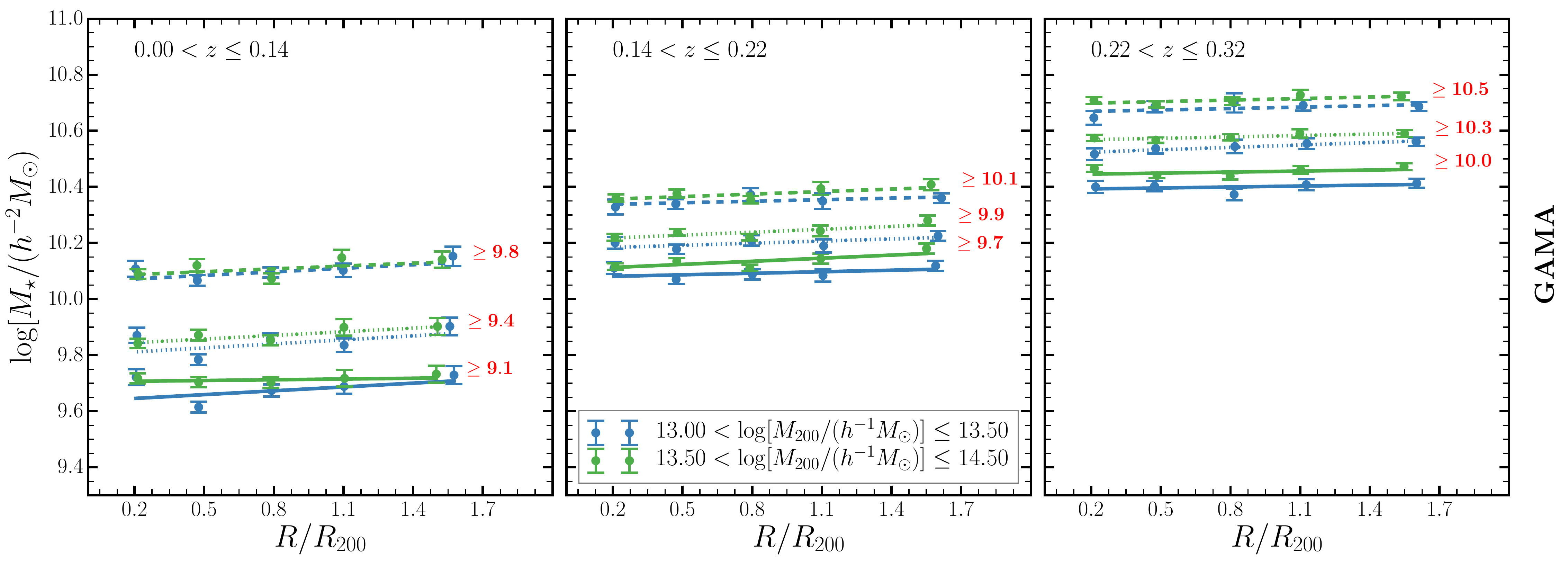}   
   \caption{Effect of stellar mass completeness limits on \gama\ data at three different redshift ranges, namely $0<z\leqslant0.14$ (left panel),
   $0.14<z\leqslant0.22$ (mid-panel) and $0.22<z\leqslant0.32$ (right panel) in the two most massive halo mass bins.
   The red texts shown alongside the mass segregation trends are the corresponding stellar mass limits applied to each sub-sample.}
\label{fig:vollimit_effect}
\end{figure*}

Here, we investigate the effect of stellar mass completeness limits in the mass segregation profiles of the \gama\ data. 
As discussed in Section~\ref{sec:vollimit}, we make an attempt to use a volume complete sample throughout our analysis. 
For this we estimate a lower stellar mass limit using the running $90^{\text{th}}$ percentile of the stellar mass distribution at all redshifts for the \gama\ data.
But the veracity of the choice of $90^{\text{th}}$ percentile can be questioned.   
Therefore, in Figure~\ref{fig:vollimit_effect} we show the impact of our choice of the percentiles 
in the mass segregation trends in \gama\ data in all the three redshift ranges 
namely $0<z\leqslant0.14$ (left panel), $0.14<z\leqslant0.22$ (mid-panel) and $0.22<z\leqslant0.32$ (right panel).
The green and blue lines represent the two most massive halo mass bins. 
We show only the most massive halo mass bins as they are the only cases where we still have enough galaxies left in each radial
and redshift bins even for an extreme choice of the stellar mass limits.
The solid, dotted and dashed lines show mass segregation trends in the \gama\ data with 
stellar mass limit estimated at the $90^{\text{th}}$, $75^{\text{th}}$ and $50^{\text{th}}$ percentiles of the distribution respectively.
We find that the slopes of all the shown trends are $\lesssim0.5$ and consistent with zero gradient given the uncertainties.
It suggests that our adopted set of stellar mass limits as a function of redshift provides a reasonable compromise between 
sample completeness and sample size.

\bsp	\label{lastpage}
\end{document}